\documentclass[twocolumn]{IEEEtran}
\IEEEoverridecommandlockouts
\usepackage{multirow}
\usepackage{tabu}
\usepackage{dirtytalk}
\usepackage{comment}
\usepackage{url}
\usepackage{mathtools}
\usepackage{array}
\usepackage{float}
\newcolumntype{P}[1]{>{\centering\arraybackslash}p{#1}}
\newcolumntype{M}[1]{>{\centering\arraybackslash}m{#1}}
\newcolumntype{B}[1]{>{\centering\arraybackslash}b{#1}}
\usepackage{amsmath,amssymb,amsfonts}
\usepackage[linesnumbered,ruled,vlined,commentsnumbered]{algorithm2e}
\usepackage{graphicx}
\usepackage{textcomp}
\usepackage{gensymb}
\usepackage{tikz}
\usepackage{xcolor}
\usepackage{caption}
\usepackage{subcaption}
\usepackage{cite}
\usepackage[utf8]{inputenc}
\usepackage[english]{babel}
\usepackage{colortbl}
\definecolor{LightGray}{gray}{0.85}
\usepackage{amsthm}

\usepackage{setspace}
\usepackage{footmisc}

\usepackage{dsfont}
\usepackage[separate-uncertainty = true,multi-part-units = repeat, load-configurations = abbreviations]{siunitx}
\usepackage{siunitx}
\DeclareMathOperator*{\argmax}{arg\,max}

\begin{document}
\title{Cooperative Multigroup Broadcast 360\degree\ Video Delivery Network: A Hierarchical Federated Deep Reinforcement Learning Approach}
\author{
\IEEEauthorblockN{Fenghe Hu, ~\IEEEmembership{Student Member, ~IEEE,}}
\IEEEauthorblockN{Yansha Deng, ~\IEEEmembership{Member, ~IEEE,}}\\
\IEEEauthorblockN{A. Hamid Aghvami, ~\IEEEmembership{Fellow, ~IEEE,}}

\thanks{F. Hu, Y. Deng, and A. H. Aghvami are with  King's College London, UK (E-mail:{fenghe.hu, yansha.deng, hamid.aghvami@kcl.ac.uk})(Corresponding author: Yansha Deng).}
}

\maketitle

\begin{abstract}
    With the stringent requirement of receiving video from the unmanned aerial vehicle (UAV) from anywhere in the stadium of sports events and the significant-high per-cell throughput for video transmission to virtual reality (VR) users, a promising solution is a cell-free multi-group broadcast (CF-MB) network with cooperative reception and broadcast access-points (AP). To explore the benefit of broadcasting user-correlated decode-dependent video resources to spatially correlated VR users, the network should dynamically schedule the video and cluster APs into virtual cells for a different group of VR users with overlapped video requests.
    \textcolor{black}{By decomposing} the problem into scheduling and association sub-problems, we first introduce the conventional non-learning-based scheduling and association algorithms, and a centralized deep reinforcement learning (DRL) association approach based on the rainbow agent with a convolutional neural network (CNN) to generate decisions from observation. 
    To reduce its complexity, we then decompose the association problem into multiple sub-problems, resulting in a networked-distributed Partially Observable Markov decision process (ND-POMDP). To solve it, we propose a multi-agent deep DRL algorithm. 
    To jointly solve the coupled association and scheduling problems, we further develop a hierarchical federated DRL algorithm with scheduler as meta-controller, and association as the controller.
    Our simulation results show that our CF-MB network can effectively handle real-time video transmission from UAVs to VR users. 
    Our proposed learning architecture is effective and scalable for a high-dimensional cooperative association problem with increasing APs and VR users. 
    Also, our proposed algorithms outperform non-learning based methods with significant performance improvement. 
\end{abstract}

\section{Introduction}
Unmanned aerial vehicle (UAV) systems bring fast and easy accessibility of aerial video capture into our daily life. Although the existing WiFi or Long-Term-Evolution (LTE) technologies can support low-resolution video transmission for flight control, they are not suitable for applications, where many audiences need simultaneous streaming high-resolution videos from UAVs for enhancing viewing experience with virtual reality contents in large sports events. A typical solution is the True View Technology for large sports events introduced by Intel \cite{Intel}. With the help of a large camera array distributed around the stadium and on UAVs, the overall scenes can be rebuilt in real-time with a huge amount of viewpoints. The captured video from diverse angles is processed into a volumetric video set, which contains real-time content for virtual reality (VR) video resources from different viewpoints. This enhances the audiences' viewing experience by immersing audiences in their selected environment with head-mounted displays (HMD). To realise the full vision of event enhancing VR service, a wireless network is needed to receive, process and transmit the captured $360\degree$ VR video from multiple UAVs to massive VR users. However, as shown by Qualcomm \cite{Hu2020,Qualcom}, the overall capacity requirement for such service from network to VR users can reach $22 Tbps/km^2$ level, which can't be satisfied with existing wireless technologies. Also, this service requires seamless real-time responses to VR users' viewpoint selections, and the newly generated video frames should be successfully transmitted and decoded without noticeable jitter or delay \cite{Hu2020,Chen2020}. 

Existing researches on VR video transmission mainly focus on reducing the transmission delay via caching and wireless resource allocation for pre-stored video resources \cite{chen2019deep,Yang2018a,Sukhmani2018,Chen2019,Hou2019}. In \cite{chen2019deep,Yang2018a,Sukhmani2018}, the authors designed a caching algorithm to reduce the transmission delay of VR video resources from the UAVs or cloud server to the VR users with the support of the edge server. By periodically re-arranging the video resource held at the edge server, the requested video resource can be directly transmitted to the VR users from the edge server without fetching from the UAVs in real-time to save the overall delay. In \cite{Chen2019}, the authors optimized the resource allocation for VR video transmission under the consideration of data correlation. With the human factor in the loop, the authors \cite{Hou2019} extended \cite{Chen2019} by integrating the prediction of VR users' motions prediction into allocation algorithm and reduce the overall delay of the video resource transmission. However, \cite{Chen2020,Perfecto2018} assumed pre-stored independent VR video resource in the form of chunk or image without considering video increment decoding schemes. In \cite{Yang2018a}, a scheduling algorithm was applied to manage the processing and transmission of correlated tasks in VR. However, their models are not for real-time VR video capture and transmission.

To satisfy the critical requirement of transmitting real-time VR video from UAVs to a large number of VR users, the broadcasting technique is shown to be a promising solution \cite{Perfecto2018}, especially for the scenario with highly correlated requests. By discretizing the video resources into smaller units, namely, tiles \cite{3GPP2018}, the correlated tiles can be broadcasted to all VR users requesting these tiles. This can largely reduce the bandwidth requirement. However, the performance of the broadcast system in a large area network is heavily limited by inter-cell interference, especially for cell-edge VR users. To cope with this challenge, one possible solution is cooperative transmission, which has been proposed in \cite{Ge2017,Ngo2017,Buzzi2017}. To facilitate a wide range of cooperation among a large number of distributed access points (AP), the authors in \cite{Ngo2017} proposed a cell-free (CF) multi-input-multi-output (MIMO) network by connecting APs to a central server via high-speed backhaul links. This concept is further extended to a user-centric CF-MIMO network, where the APs are clustered into different groups that can serve multiple groups of users simultaneously \cite{Buzzi2017}. However, the association problem in such a network is complex to solve due to the exponential increase of the complexity with the number of cooperative APs \cite{Buzzi2017}. Besides, the environment information is high-dimensional with a large number of VR users and cooperative APs. Luckily, deep reinforcement learning (DRL) has been shown useful in solving high dimensional communication problems in complex environments \cite{Chen2019,Perfecto2018,Cui2019}, but its scalability is still an issue for multiple-agent large-scale networks. 

Motivated by the above, in this paper, a CF broadcast network is proposed to jointly stream the VR video resources from UAVs and broadcast to the target VR user groups with spatial and content correlation. In this network, there are two challenging problems to solve in real-time: 1) a scheduler to arrange the transmission and re-transmission of VR video resources; 2) an association algorithm to dynamically re-group APs to connect UAVs with each VR user group and reduce interference. Importantly, the scheduling and association stages occur sequentially. More specially, the scheduler first decides the tiles to be transmitted. The optimal association then group APs based on UAV and VR users' positions to avoid high inter-cell interference. This calls for a joint design of scheduling and association algorithms. Our contributions are summarized as follows:

\begin{itemize}
    \item \textcolor{black}{We first propose a decode-forward (DF) CF-MB network for VR video resource transmission with \emph{UAV-APs} uplink from UAV camera to APs group, and \emph{APs-VR} downlink from APs group to users. We also define our VR video resource via tiles, and QoE metric via the viewpoint-peak-signal-noise-ratio (V-PSNR) based on the number of successfully decoded tiles at the VR users' sides. Then, we formulate our optimization problem as a semi-Markov-decision-process (semi-MDP).}
    \item \textcolor{black}{We then highlight the limitation of conventional centralized learning algorithms where the complexity of the problem increases exponentially with an increasing number of participating APs. To cope with this challenge, we first formulate the association part of the optimization problem as a networked partially observable Markov-decision-process (ND-POMDP) via mean-field theorem. In this way, we decompose the association problem into multiple subproblems, which are networked coordinated. We then propose a distributed multi-agent DRL approach with the help of federated to stabilize and accelerate the learning. Our results highlight that our distributed algorithm can efficiently solve the association optimization problem with decent scalability. The existing learning algorithms from existing works fail to capture the scalability problem and are only capable to deal with several access-points (APs).}
    \item \textcolor{black}{To jointly optimise the interplay between the scheduling and the association, we propose a hierarchical DRL architecture with a centralized scheduler and distributed association to jointly optimize the V-PSNR. Our results show that the hierarchy learning structure can effectively handle the complex optimization problem with sequential decisions.}
\end{itemize}
The remainder of this paper is organized as follows. Section II illustrates the communication model and video decoding model. In Section III, we define our optimization target by defining viewpoint peak signal-noise ratio (V-PSNR) as the QoE metric. We propose conventional methods for scheduling and association separately. In Section IV, we first propose our centralized DRL algorithm for the association problem. We introduce ND-POMDP problem, which is then solved in a federated multi-agent association setting. Then, in Section V, we apply the hierarchical learning method to capture both scheduling and association sub-problems. The numerical results are presented in Section VI. Finally, we conclude the paper in Section VII.

\begin{figure*}[t]
     \centering
     \begin{subfigure}[b]{0.28\textwidth}
         \centering
         \includegraphics[width=\textwidth]{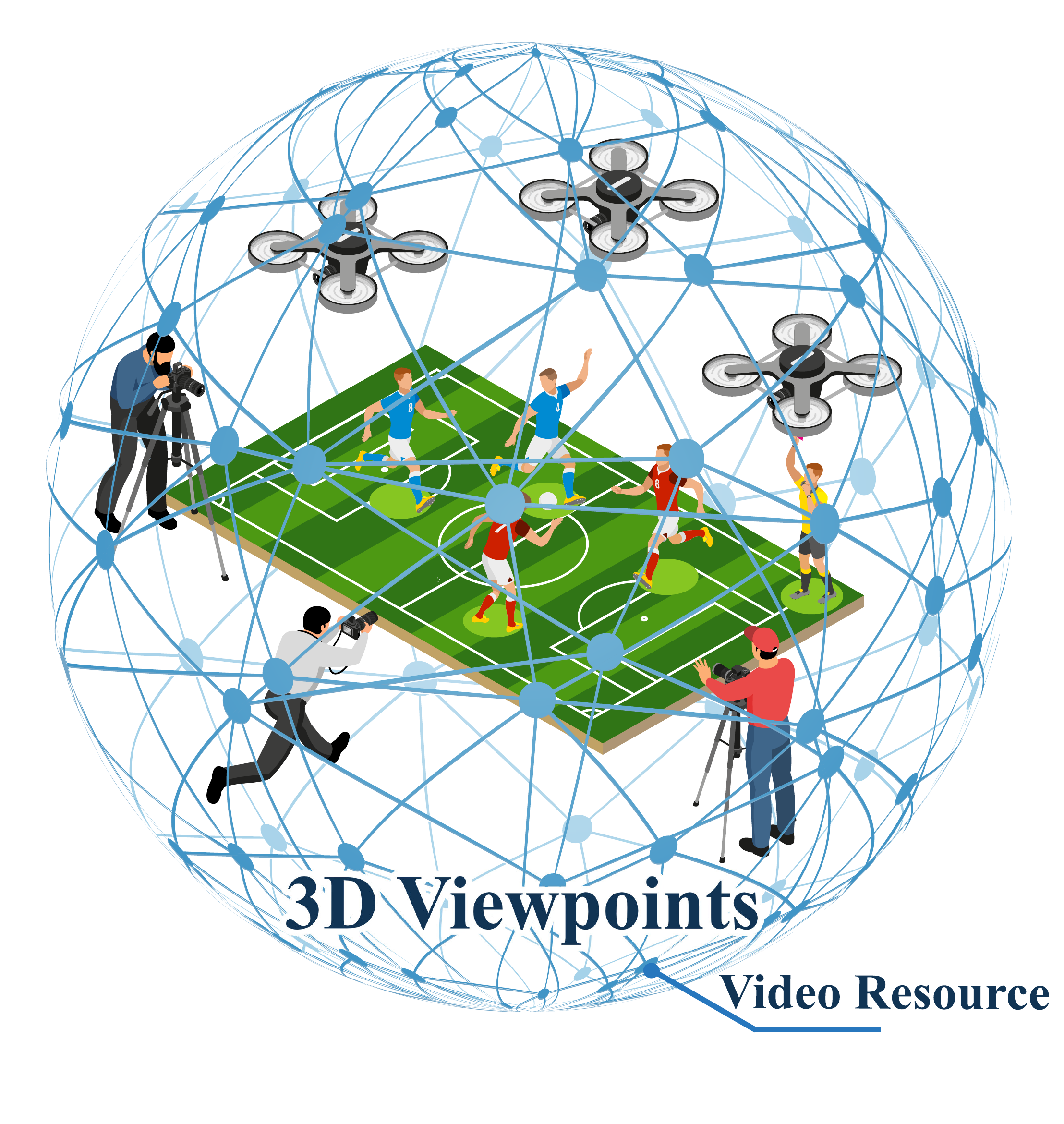}
         \caption{Event scenario with viewpoints captured by multiple UAVs}
         \label{fig:video_model_capture}
     \end{subfigure}
     \hfill
     \begin{subfigure}[b]{0.28\textwidth}
         \centering
         \includegraphics[width=\textwidth]{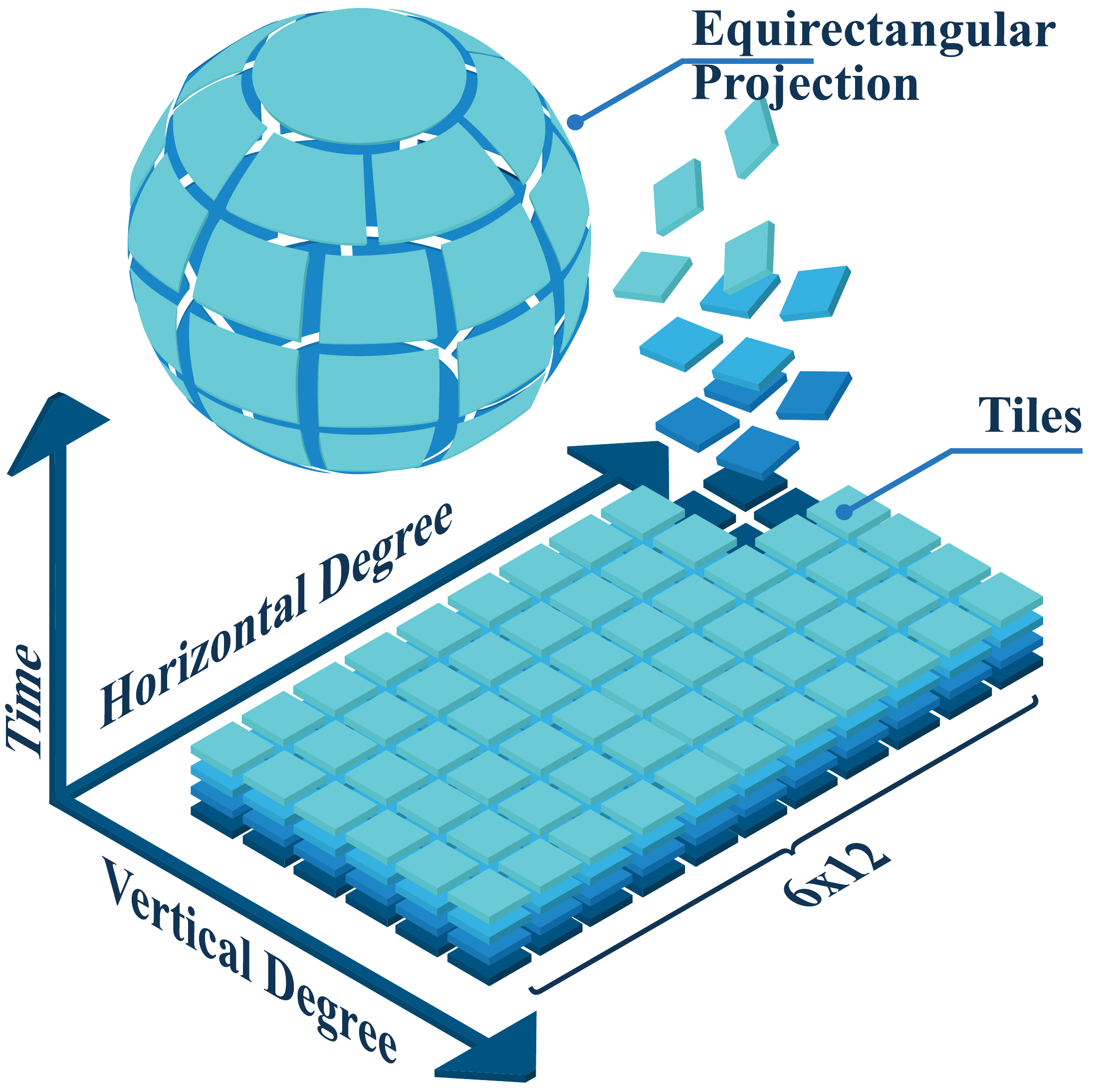}
         \caption{Tiled-based video mapping from $360\degree$ to 2D}
         \label{fig:video_model_res}
     \end{subfigure}
     \hfill
     \begin{subfigure}[b]{0.41\textwidth}
         \centering
         \includegraphics[width=\textwidth]{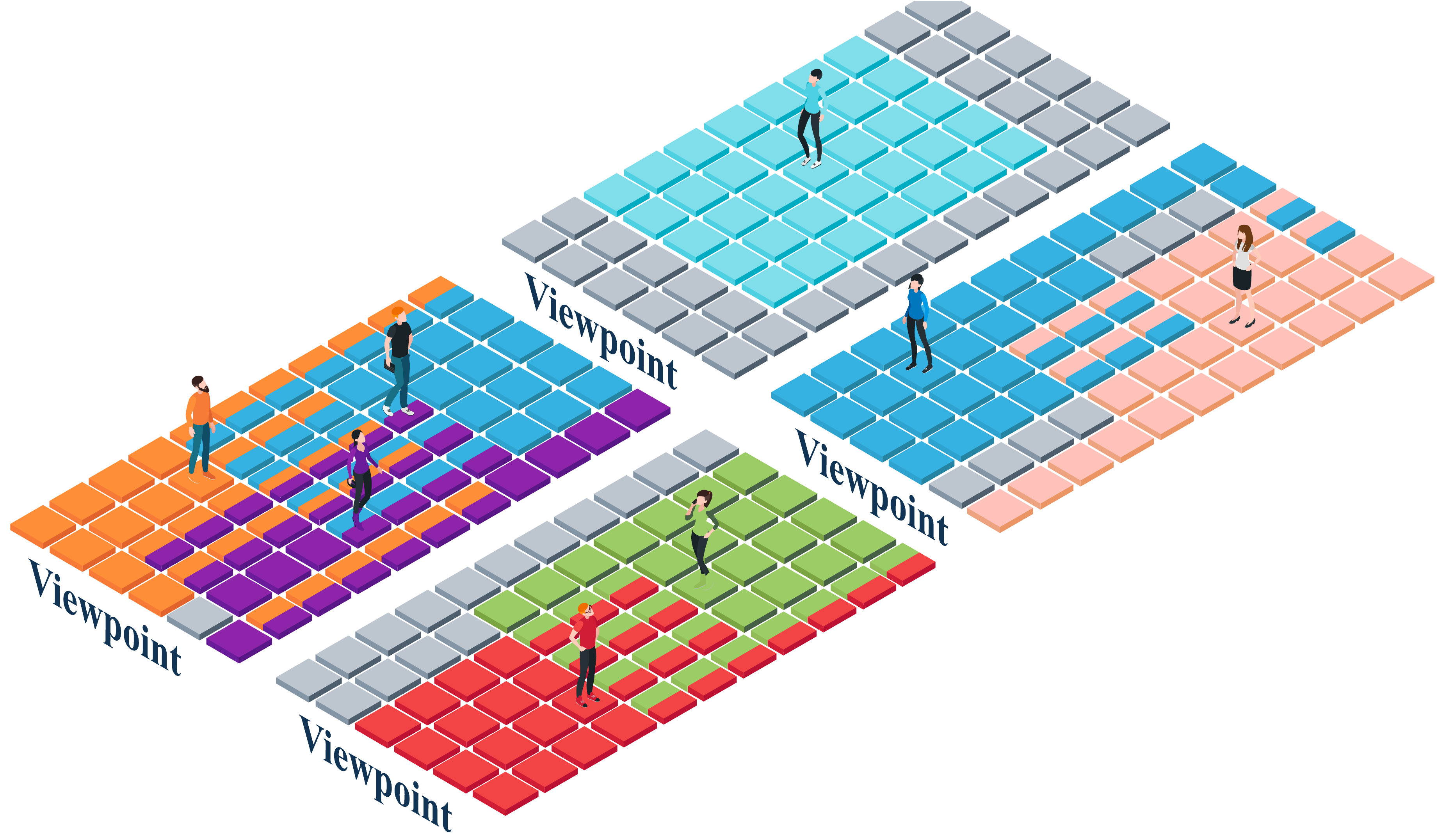}
         \caption{User correlation in tiles}
         \label{fig:video_model_corre}
     \end{subfigure}
        \caption{Illustration of scenario, tiled-based video model and corresponding VR user correlation.}
        \label{fig:video_model}
\end{figure*}

\section{System Model}
As illustrated in Fig. \ref{fig:video_model_capture} and Fig. \ref{fig:system_model}, we consider a cell-free multi-group broadcast (CF-MB) network for $360\degree$ video transmission in a large sports event. This CF-MB network is composed of 1) a set of APs $\mathcal{B}$, which are located in the grid; 2) a central server, which connects all APs through backhaul optical links; 3) a set of randomly located camera UAVs $\mathcal{U}$, where each UAV provides the video resource from their orientation; and 4) a set of VR $\mathcal{V}$ users, whose locations follow Poisson cluster process (PCP) with $|\mathcal{U}|$ clusters \cite{saha2017poisson}. \textcolor{black}{We consider the distribution of VR users follows PCP because the user density is highly correlated in the hot-spot area in large events or large-scale networks \cite{Shang2020Machine}, such as stores and certain booths.
It is also noted that the VR users' video requests can correlate to their locations in our considered scenario. Thus, PCP is more realistic for a user-centric correlated scenario than other uniform distributions.}

\textcolor{black}{Each VR user requests its video resource from a UAV based on their field-of-view (FOV) selection. As shown in Fig.\ref{fig:video_model_res}, VR users first select their interested FoV provided by multiple UAVs. Then, as shown in Fig.\ref{fig:video_model_corre}, VR users request the tiles from $360\degree$ video provided by the selected UAV based on their FOV selection. Also, as shown in Fig. \ref{fig:video_model_corre}, the overlapped FoV results in the correlation among VR users' requests \cite{Perfecto2018}. 
We assume that VR users in each PCP cluster request video resources from the same UAV but different tiles, while the clusters can be overlapped or disjointed.} All nodes are located inside the serving area of the plane $\mathbb{R}^2$, and remain spatially static in each group-of-picture (GOP) once deployed. The video resource is captured by UAVs, processed by the central server, and transmitted to VR users via the CF-MB network on request. In short, the CF-MB network acts as a decoded-forward (DF) relay, which receives the video from the UAV and broadcasts the processed video to target the VR user group based on VR users' requests. However, the resource requests are small packets in tens of bytes level, whereas the video data's size is usually in GB level. Due to the significant different traffic characteristics of video data and VR requests, we focus on the \emph{UAV-APs} uplink from UAVs to the APs in the CF-MB network, and the \emph{APs-VR} downlink from the APs in the CF-MB network to VR users for video resource in this paper.

\subsection{Transmission Channel Model}
To capture the different channel characteristics between APs, UAVs and VR user groups, we consider different channel models for the \emph{UAV-APs} uplink and the \emph{APs-VR} downlink, respectively. 
The \emph{UAV-APs} uplink from UAV to APs and APs-UAV downlink from APs to VR user group occupy $B_{\text{UL}}$ and $B_{\text{DL}}$ bandwidth, respectively. We also assume that a perfect channel state information (CSI) is available at the APs. We assume that both channels follow the block fading assumption, where the channel remains constant on a time-frequency coherence block \cite{Karimi2018}.

\subsubsection{UAV-APs Uplink}
The \emph{UAV-APs} uplink between a UAV and a AP group forms a virtual single-input-multi-output (SIMO) system, where multiple APs are associated to enhance the signal reception quality. Considering potential line-of-sight (LoS) and non-line-of-sight (NLoS) for low altitude flying drones, we adopt free-space path loss and Rayleigh fading to model the \emph{UAV-APs} uplink path loss model as
\begin{equation}
    h_{u,b}=
        \begin{cases}
        (\frac{4\pi d_{u,b} f^{\text{UL}}_\text{c}}{c})^{\alpha_{\text{UL}}} \eta_{\text{LoS}} \beta_{u,b} , & P_{\text{LoS}}^{u,b}\\
        (\frac{4\pi d_{u,b} f^{\text{UL}}_\text{c}}{c})^{\alpha_{\text{UL}}} \eta_{\text{NLoS}} \beta_{u,b} , & P^{u,b}_\text{NLoS} = 1-P^{u,b}_\text{LoS}
        \end{cases},
        \label{los}
\end{equation}
where $\theta_{u,b} = \frac{180}{\pi}\sin^{-1}(\frac{\text{h}_{u,b}}{d_{u,b}})$ is the elevation angle of the drone, $\text{h}_{u,b}$ represents the height of flying drone, $d_{u,b}$ denotes the distance between the $b$th AP and the $u$th UAV \cite{Mozaffari2017}, $f^{\text{UL}}_\text{c}$ is the uplink channel center frequency, $\eta_{\text{LoS}}$ and $\eta_{\text{NLoS}}$ are the excessive path loss coefficients in LoS and NLoS cases, $c$ is the light speed, and $\alpha_{\text{UL}}$ is the path loss exponent. In \eqref{los}, we adopt the LoS probability of the \emph{UAV-APs} uplink as \cite{Mozaffari2016}
\begin{equation}
    P^{u,b}_\text{LoS} = \frac{1}{1+11.95\exp(-0.14[\theta_{u,b} - 11.95])},
    \label{loschannel}
\end{equation}
where $\theta_{u,b} = \frac{180}{\pi}\times\arcsin (\frac{h_u}{d_{u,b}})$, $h_u$ is the flight height of UAV.

Based on \eqref{los} and \eqref{loschannel}, the combined channel between the $b$th AP and the $u$th UAV can be expressed as
\begin{equation}
    h_{u,b}= [P^{u,b}_\text{LoS}\eta_{\text{LoS}} + P^{u,b}_\text{NLoS}\eta_{\text{NLoS}}](\frac{4\pi d_{u,b} f^{\text{UL}}_\text{c}}{c})^{\alpha_{\text{UL}}}\beta_{u,b},
\end{equation}
where $P^{u,b}_\text{LoS}$ is given in \eqref{loschannel}.

\subsubsection{AP-VR Uplink}
We consider Rayleigh fading for multi-input-single-output (MISO) transmission between each AP group and VR user \cite{Ngo2017}. The channel between the $b$th AP and $v$th VR user is represented as 
\begin{equation}
    h_{b, v}= d_{b, v}^{-\alpha_{\text{DL}}} \beta_{b, v},
    \label{channel_model}
\end{equation}
where $d_{b, v}$ represents the distance between the $b$th AP and the $v$th VR user, $\alpha_{\text{DL}}$ represents the AP-VR uplink path loss exponent, and $\beta_{b,v}$ denotes the Rayleigh small-scale fading.

\subsection{Tile Transmission Data Rate}
\textcolor{black}{From the perspective of the CF-MB network, the network is operating in frequency-division-duplex (FDD) mode with the transmission of \emph{UAV-APs} uplink and \emph{APs-VR} downlink at the same time over different frequency bands }. We assume that all UAV and VR users are equipped with one antenna. Each AP is equipped with two antennas where one antenna for \emph{UAV-APs} uplink, one for \emph{APs-VR} downlink. The tile transmission model can be seen as a DF relay system, where \emph{UAV-APs} uplink is SIMO transmission and \emph{APs-VR} downlink is MISO transmission.

\subsubsection{Data Rate of UAV-APs Uplink}
For the SIMO transmission of \emph{UAV-APs} uplink from single UAV to multiple cooperative APs, we adopt the maximum-ratio combining (MRC) technique to realise the multiple reception gain. The received signal $\gamma_{u^*, b}$ from scheduled the $u^*$th UAV to the $b$th AP within associated APs group $\mathcal{B}^{u^*}_t$  at time $t$ can be expressed as
\begin{equation}
y_{u^*, b}= \underbrace{\vphantom{\sum\limits_{u',u'\neq u^*}^{\mathcal{U}}  h_{u', b} s_{u'}} h_{u^*,b} s_{u^*}}_{\text{Desired signal}}
+ \underbrace{ \sum\limits_{u'\in\mathcal{U}_t\backslash u^*}^{\mathcal{U}} h_{u', b} s_{u'}}_{\text{Interference from Other UAVs'}}
+ \underbrace{\vphantom{\sum\limits_{u',u'\neq u^*}^{\mathcal{U}}  h_{u', b} s_{u'}} n_0}_{\text{Noise}},
\end{equation}
where $h_{u,b}$ denotes the channel vector from the $u$th UAV to the $b$th AP, $\mathcal{U}_t$ is the current scheduled UAV, $h_{u', b}$ is the interference channel from other interfering UAVs, $s_{u}$ is the signal transmitted by the $u$th UAV, $N_0 \sim \mathcal{CN}(0,I_{N})$ represents the Gaussian white noise. 
Then, the signal after MRC can be expressed as
\begin{equation}
    \gamma_{u^*, \mathcal{B}^u_t} = \sum_{b\in\mathcal{B}^u_t} w_{b} y_{u^*, b},
\end{equation}
where $w_b$ is a general weighted MRC scheme with weight $w_{b} = h_{u^*, b}^H/||\mathbf{h}_{u^*, \mathcal{B}^u_t}||_\text{F}, \ b\in\mathcal{B}^u_t$, $||\cdot||_\text{F}$ represents Frobenius norm, and $\mathbf{h}_{u^*, \mathcal{B}^u_t} = [h_{u^*, b_0},...,h_{u^*, b_{|\mathcal{B}^u_t|}}]$ is a $|\mathcal{B}^u_t| \times 1$ channel vector from target the $u^*$th UAV to a corresponding APs group $\mathcal{B}^u_t$ \cite{Jindal2011}.

Thus, the received SINR for tile upload from the $u$th UAV to accesspoint group $\mathcal{B}^k_t$ at time $t$ can be expressed as
\begin{equation}
\gamma_{u^*, \mathcal{B}^u_t}= \frac{\sum\limits_{b\in\mathcal{B}^u_t} p_{u} |w_{b} h_{u, b}|^2}{\sum\limits_{b\in\mathcal{B}^u_t}\sum\limits_{u'\in\mathcal{U}\backslash u}^{\mathcal{U}} p_{u'} |w_{b} h_{u', b}|^2 + \sum\limits_{b\in\mathcal{B}^u_t}|w_{b}|^2 \sigma^2}.
\end{equation}

Due to the flat-fading in each broadcast slot, the received data capacity $D_{{u^*}, \mathcal{B}^{u^*}_t} (t)$ during resource block at the group of APs $\mathcal{B}^{u^*}_t$ from the $u^*$th UAV is given by
\begin{equation}
    D_{u^*, \mathcal{B}^{u^*}_t} = T_\text{b}B_{\text{UL}} \log_2(1 + \gamma_{u^*, \mathcal{B}^{u^*}_t}).
\end{equation}

\subsubsection{Data Rate of the APs-VR Uplink}
In \emph{APs-VR} uplink, the APs form virtual-cells to jointly broadcast the tiles to corresponding VR user groups and enhance the broadcasting quality. As shown in Fig. \ref{fig:system_model}, the cooperative APs can enhance the signal quality in receiving from the VR users, but the inter-cluster interference limits the overall performance. To realise the gain of jointly broadcasting and to improve the worst VR user's performance, we adopt linear sum maximum precoding \cite{Joung2015}. With perfect channel state information (CSI), the precoding matrix in the $b$th AP can be given by
\begin{equation}
    w_{b} = \alpha_b \sum\limits_{v}^{\mathcal{V}^k_t} \frac{h_{b, v}^H}{||h_{b, v}||^2},\ b\in\mathcal{B}^k_t,
    \label{apsvrchannel}
\end{equation}
where $\alpha_b$ is the normalize factor to ensure $||w_b||^2_\text{F}=1$.

Based on \eqref{apsvrchannel}, the signal received at the selected $v^*$th VR user ($v^*\in\mathcal{V}^u_t$) from the $b$th AP can be expressed as
\begin{equation}
\begin{aligned}
y_{\mathcal{B}^u_t, v^*} =& \sum\limits_{b\in\mathcal{B}^u_t} h_{b, v^*}w_{b} s_{b} + 
    \sum\limits_{b'\in\mathcal{B}\backslash \mathcal{B}^u_t}^{\mathcal{B}} h_{b', v^*}w_{b'} s_{b'} + n_{v^*} \\
=& \underbrace{\sum\limits_{b\in\mathcal{B}^u_t} h_{b, v^*} \sum\limits_{v}^{\mathcal{V}^u_t}\alpha_{b}\frac{h_{b, v}^H}{||h_{b, v}||^2} s_{b}}_{\text{Desired Signal}} \\
&+  \underbrace{\sum\limits_{\mathcal{B}^n_t\in\mathcal{B}\backslash \mathcal{B}^u_t}^{\mathcal{B}} \sum\limits_{b'}^{\mathcal{B}^n_t} h_{b', v^*} \sum\limits_{v'}^{\mathcal{V}^{n}_t}\alpha_{b'}\frac{h_{b', v'}^H}{||h_{b', v'}||^2} s_{b'}}_{\text{Inter-group Interference}} + \underbrace{\vphantom{\sum\limits_{\mathcal{B}^n_t\in\mathcal{B}\backslash \mathcal{B}^u_t}^{\mathcal{B}} \sum\limits_{b'}^{\mathcal{B}^n_t} h_{b', v^*} \sum\limits_{v'}^{\mathcal{V}^{n}_t}\alpha_{b'} \frac{h_{b', v'}^H}{||h_{b', v'}||^2} s_{b'}}n_{v^*}}_{\text{Noise}},
\end{aligned}
\label{apsvr}
\end{equation}
where $w_b$ denotes the precoding matrix for the $b$th AP in group $\mathcal{B}^u_t$ at time $t$, and $h_{b, v}$ is the path loss for the channel between the $b$th AP and the $v$th VR user at time $t$. 

Based on \eqref{apsvr}, the SINR from the $b$th AP in APs group $\mathcal{B}_t^u$ to $v^*$th VR user in user group $\mathcal{V}^u_t$ at time $t$ can be expressed as
\begin{equation}
\gamma_{\mathcal{B}^u_t, v^*}=\frac{\sum\limits_{b\in\mathcal{B}^u_t} p_{b} |h_{b, v^*} w_{b}|^2}{\sum\limits_{b'\in\mathcal{B}\backslash \mathcal{B}^u_t}^{\mathcal{B}} p_{b'} |h_{b', v^*}w_{b'}|^2 + \sigma^2}.
\end{equation}

Under given SINR, the received data $D_{\mathcal{B}^u_t, v}$ in one broadcast slot $T_\text{b}$ from the APs group $\mathcal{B}^u_t$ to $v^*$th VR user can be calculated by the minimum ergodic rate within the broadcast group
\begin{equation}
    D_{\mathcal{B}^u_t, v} = T_\text{b}B_c^{DL}\log_2(1 + \gamma_{\mathcal{B}^u_t, v^*}).
\end{equation}

\subsection{Tiled-based Video Resource Model}
Tile-based VR video transmission can effectively support the broadcasting of video resources \cite{3GPP2018}. It splits the captured video resource into small tiles, which can be decoded individually. By exploring the nature of video codec, the tiled-based video transmission is introduced for VR video transmission, where the tiles in the same location can be decoded individually \cite{3GPP2018}.
As shown in Fig. \ref{fig:video_model_capture}, each UAV records a $360\degree$ video stream with on-broad camera, which is converted and transmitted in 2D video format via Equirectangular projection. As shown in Fig. \ref{fig:video_model_res}, \textcolor{black}{we define that each tile contains for $30\degree\times 30\degree$ square part in $180\degree\times 360\degree$ video, which is full-view from a certain viewpoint} \cite{Zink2019a}. Thus, each UAV provides $6\times 12$ tiles, which is shown in Fig. \ref{fig:video_model_res}. The size of one tile is defined as $\mu M_\text{T}$ bits, where $\mu$ is the compression rate. The compression rate is decided by tile type, which is explained later. We also denote the overall tile set as $\mathcal{J}$, which is provided by the set of UAVs $\mathcal{U}$. 

As shown in Fig. \ref{fig:system_model}, a set of new tiles from newly captured video frames are generated every $T_\text{f}$ time, i.e. at frame rate $1/T_\text{f}$. In our scenario, we assume that all video frames from different UAVs are captured and encoded in the same frame rate.

\subsection{Tiles Requests, Receiving and Decoding Model}
To describe the content request in each VR user via tiles, \textcolor{black}{we highlight that the field-of-view (FoV) of human is defined as $150\degree\times 210\degree$ \cite{3GPP2018}. Thus, $v$th VR user requests $5\times 7$ tiles, denoted by $\mathcal{J}^v_t,\ v\in\mathcal{V}$ at time $t$ ($|\mathcal{J}_t^v| = 5\times 7 = 35$).} The actual number of tiles can vary based on different positions of viewpoints in $360\degree$ space which requires tiles follow the rule of 3D-2D projection \cite{Xu2019} (as shown in Fig. \ref{fig:video_model_corre}), and viewpoints are randomly generated. In this way, the group of VR users who requests the same tile $j$, can be served via the broadcast channel at the same time. We denote this group of VR users as $\mathcal{V}_j$. The highly correlated tile requests in our scenario highlight the potential benefit of broadcasting overlapping tiles.

To transmit the requested tiles to corresponding VR users, the tiles are delivered via the aforementioned DF network transmission, since the success of tile transmission will only occur when both the \emph{UAV-APs} uplink and \emph{APs-VR} downlink success. The successful transmission of $j$th tile can be written as the combination of successful transmission in \emph{UAV-APs} uplink and \emph{APs-VR} downlink as
\begin{equation}
    \mathds{1}[D_{u, v} \geq \mu M_\text{T}] = \mathds{1}[D_{u, \mathcal{B}^u_t} \geq \mu M_\text{T}] \land \mathds{1}[D_{\mathcal{B}^u_t, v} \geq \mu M_\text{T}],
    \label{Capacity_prob}
\end{equation}
where $\mu M_\text{T}$ is the size of tile to be transmitted, $\mathds{1}[x]=1$ as $x$ is true, $\mathds{1}[x]=0$, otherwise. $\land$ is logical and operation. $\mathds{1}[x]\land\mathds{1}[x]=1$ as $x$ and $y$ is true, $\mathds{1}[x]\land\mathds{1}[x]=0$.

\begin{figure*}[t]
    \centering
    \includegraphics[width = 0.8\textwidth]{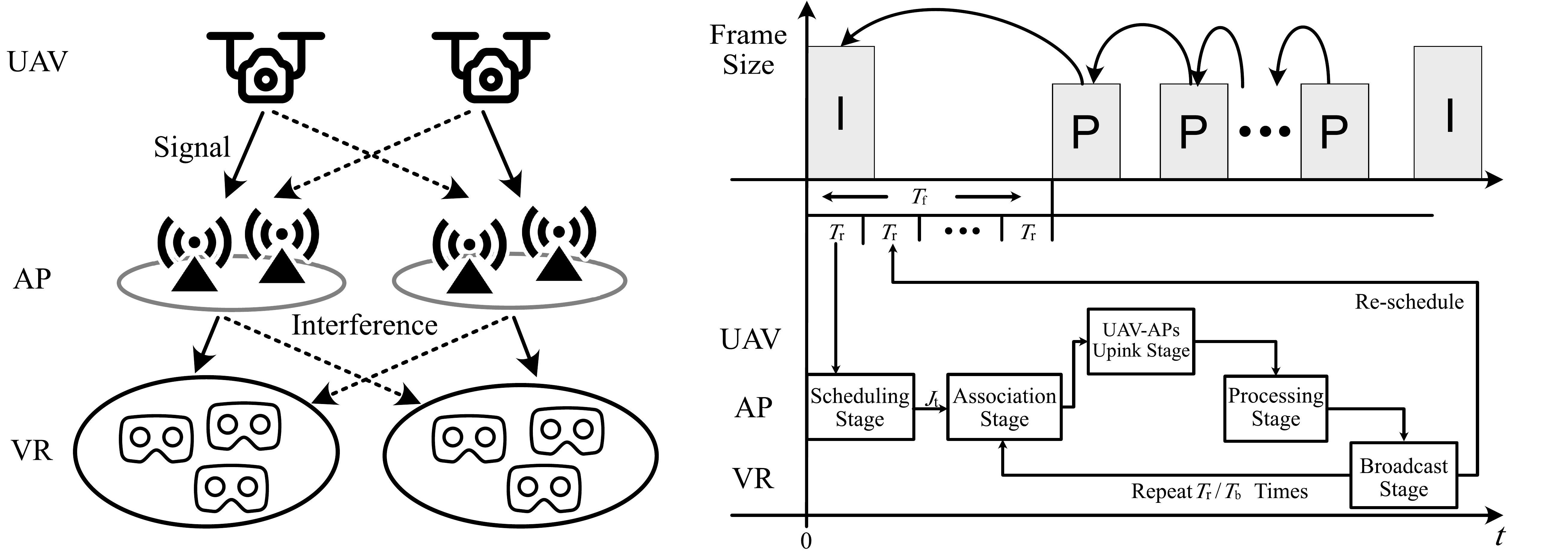}
    \caption{Communication stages and video tiles decoding relationship for considering system.}
    \label{fig:system_model}
\end{figure*}

After receiving the tiles, the tiles need to be decoded dependently with frame decoding scheme, as shown in Fig. \ref{fig:system_model}, since frames are encoded incrementally within group-of-pictures (GOP) to reduce the overall data rate. For the low-latency video encoding scheme, we only consider two typical kinds of the frame inside one GOP --- intra-coded frame (I Frame), and predicted-coded frame (P frame). The I frame can be decoded individually, whereas the P frame requires the same location's frame or tile in previous time instance to decode \cite{h265}. With such a dependent encoding scheme, the overall channel capacity required for video transmission can be saved. Thus, one tile can be successfully decoded only when the previous tiles are successfully decoded, whose set is denoted as $\mathbf{J}^v_t$ in the $v$th VR user at time $t$.

\subsection{Network Transmission Procedure}
We show the transmission procedure for tiles from UAV to VR users in this subsection. Recap that in our considered CF-MB network, the network performs as a DF relay system to support the tile $j$ transmission from $u$th UAV to VR user group $\mathcal{V}_j$ via APs group $\mathcal{B}^u_t$. For time-frequency resources, we adopt a time-division duplex, which is assumed in many massive MIMO works~\cite{flordelis2018massive}. 

As shown in Fig. \ref{fig:system_model}, each frame with a duration of $T_\text{f}$ is divided into $T_\text{f}/T_\text{r}$ re-scheduling slots, where $T_\text{r}$ is the length of each re-scheduling slot. Each re-scheduling slot is further divided into $T_\text{r}/T_\text{b}$ broadcast slots, where $T_\text{b}$ is the length for each broadcast slot. As shown in Fig. \ref{fig:system_model}, there are 5 stages in each re-scheduling slot of the network transmission procedure, which are scheduling stage, association stage, \emph{UAV-APs} uplink transmission stage, processing stage, and \emph{APs-VR} downlink transmission stage. In the scheduling stage, the network first decides the priority of tiles in each UAV based on VR users' requests. Then, the $\lfloor T_\text{r}/T_\text{b}\rfloor$ tiles with highest priority in each UAV is picked for transmission within $T_\text{r}$. 

In the association stage, each AP selects one UAV and corresponding VR users to serve inside each broadcast slot $T_\text{r}$, since the single-antenna UAV is capable of transmitting one tile at the same time. We denoted the target VR users whose requested $j$th tile is transmitted at time $t$ forms the VR user group $\mathcal{V}_j$ ($j\in J_t$). Then, the APs select the same UAV are clustered as a virtual cell $\mathcal{B}^u_t$ to jointly serve the same UAV and corresponding group of VR users, i.e. both \emph{UAV-APs} uplink and \emph{APs-VR} downlink. 

Then, in the \emph{UAV-APs} uplink stage, the $u$th UAV transmits the scheduled tile $j$ to it associated AP group $\mathcal{B}^u_t$, which jointly receives the signal. In the processing stage, the tile is processed at the central server, whose delay is considered as a constant value and ignored in our analysis. In the broadcast stage, the APs in virtual cell $\mathcal{B}^u_t$ jointly broadcast the tile to the VR user group requesting tile $j$, i.e. $\mathcal{V}_j$. The \emph{UAV-APs} uplink stage, processing stage, and \emph{APs-VR} downlink stage repeated until the end of $T_\text{r}$ with the same scheduling priority.

\subsection{Quality-of-experience Metric for VR Users}
\textcolor{black}{Generally, for video-based VR service, the literature defines the QoE as the break-in-presence (BIP), which describes the event when users stop responding to the virtual environment (the video frame is not delivered upon a certain threshold or the resource requirement is not satisfied) \cite{Chen2020}. However, when it comes to practical VR applications, the QoE is defined by the information loss in the video of current scenes. Sometimes, failing to transmit one of the tiles on time may not significantly influence the performance, as the difference between this failure tile and the previous tile is negligible. Thus, the BIP, which simply classifies the transmission of each frame into success and failure cases, lacks realistic meaning, detailed resolution and accuracy. Thus, we borrow the idea of Peak Signal-to-Noise (PSNR), which measures the spatial information difference between desired and received video. We then create our QoE matrix via viewport-PSNR (V-PSNR) inspired by the idea from PSNR and BIP.}

\textcolor{black}{We calculate the information differently based on the amount of successfully received and decoded tiles in each time slot with PSNR function \cite{Li2019a} and decide each tile is successfully received and decoded via BIP function \cite{Chen2020}.
By doing so, we can quantify the decoded QoE with PSNR value inside the $v$th VR user field-of-view at time $t$ using V-PSNR as}
\begin{equation}
    \text{V-PSNR}_t^v =  10\log_{10}\frac{|\mathcal{J}^v_t|}{|\mathcal{J}^v_t| - \sum_{j\in\mathcal{J}^v_t}\mathds{1}[j\in\mathbf{J}_t^v]},
    \label{totalTarget_rewrite}
\end{equation}
\textcolor{black}{
where $\mathcal{J}^v_t$ is the desired tile set, and $\mathbf{J}^v_t$ represents the successful decoded tiles at time $t$ in the $v$th VR user ($\mathbf{J}^v_t \subseteq \mathcal{J}^v_t$).
The V-PSNR value gives $20\log_{10}|\mathcal{J}^v_t|$ if all the tiles requested by the $v$th VR user are transmitted successfully.
$\mathds{1}[j\in\mathbf{J}_t^v]$ denotes the BIP function of single tile $j$ at time $t$. The BIP of tile $j$ depends on the successful transmission of its and its dependent tiles:}
\begin{equation}
    \mathds{1}[j\in\mathbf{J}^v_t]=
    \begin{cases}
    \mathds{1}[D_{u,v}\geq\mu M_\text{T}],& t< T_\text{f},\\
    \mathds{1}[D_{u,v}\geq\mu M_\text{T}]\land\underbrace{\mathds{1}[j'\in\mathbf{J}^v_t]}_{\text{Dependent tile received}},& t\geq T_\text{f}\\
    \end{cases}
    \label{decode_tile}
\end{equation}
where $\mathds{1}[D_{u,v}\geq\mu M_\text{T}]$ is given in \eqref{Capacity_prob}, $j$ and $j'$ are dependent tiles, $j$ is required to be decoded with $j'$ incrementally, i.e. $j$ depends $j'$ to decode. In each GOP, when $t<T_\text{f}$, the tile is from I frame, which can be decoded independently.

\section{Problem Formulation and Conventional Methods}
In this section, we defined and decomposed our optimization problem into scheduling and association sub-problems. We then introduce the conventional methods for each sub-problem.

\subsection{Problem Formulation}
\textcolor{black}{We aim to design an algorithm for the CF-MB network which supports the tile transmission from UAVs to VR users and enhance the QoE of VR users by dynamically adjusting the scheduling and association decisions. The system can be seen as a Markov decision process (MDP), as the system state can be fully characterised via a state $s$ without correlation with historical decisions in each time slot. Knowing that the small-scale fading is independent of historical information. We denote the set of the state as $\mathcal{S}$. Each state $s$ contains VR users' request, VR users' decoding sequence, UAVs' position, VR users' V-PSNR, \emph{UAV-APs} uplink's, \emph{APs-UAV downlink}'s channel information, and etc. The system transfers to a new state in the next time slot based on the scheduling and association decisions with the probability transfer function. However, simultaneously adjusting both decisions is complex \cite{Peng2021Multi}, especially when these two decisions are made in different time scales --- scheduling priority is updated every re-scheduling slot $T_\text{r}$, and association decision is updated every broadcast slot $T_\text{b}$.  To deal with this, in our proposed tile transmission procedure, we highlight that the scheduling and association procedures are executed successively. Scheduling is the primitive action of association. Then, it is appropriate to define the original problem as a semi-MDP with scheduler as Markov options of the association MDP, and solve it via a hierarchy architecture \cite{Sutton1999MDPs}.}

\textcolor{black}{We first define our considered problem as a semi-Markov Decision Process (semi-MDP) \cite{Sutton1999MDPs}. The decision is made via policies from both scheduling and assocation. With the scheduling policy as $\pi_{\text{s}}$ and the association policy as $\pi_{\text{a}}$, the scheduling priority and associated virtual cells $\{\mathcal{B}^u_{t_\text{b}}\}$ ($u\in\mathcal{U}$) are decided based on current state in sequence. Here, $\pi_{\text{s}}$ is a weighted mapping from the current state to the priority of tile transmission. The $T_{t_\text{r}}/T_{t_\text{b}}$ tiles with highest priority is allocated to be transmitted, whose set is denoted as $\mathcal{J}_{t_\text{r}}$. Here, $\pi_{\text{a}}$ is the distribution mapping from the current environment state and selected scheduling decisions to the selection of each UAV and corresponding VR user group. Here, the scheduling priority is considered as the primitive actions, which is followed by association decisions that persist through $T_\text{r}$. 
Then, the system state $S_{{t_\text{b}}+1}$ in $({t_\text{b}}+1)$-th broadcast slot transfers from the system state $S_{t_\text{b}}$ based on the probability transfer function. The transfer function and expected reward are defined via each state, scheduling policy, and association policy, such that it forms semi-Markov decision process (semi-MDP) problem, and the scheduling is defined as the option on the association MDP. }

\textcolor{black}{The Markov option (scheduling) is defined by a tuple of initial state set $\mathcal{I}\subseteq \mathcal{S}$, terminal condition $t=nT_\text{r}, n\in\mathbb{N}$, and policy $\pi:\mathcal{S}\rightarrow \mathcal{A}^\text{s}$ \cite{Sutton1999MDPs}. Here, we consider a constant re-schedule time. Thus, the initial state $\mathcal{I}$ contains the state when $t=nT_\text{r}, n\in\mathbb{N}$. The scheduling policy maps the state to the scheduling priority. Note that scheduling needs to run at equal or slower time-step than association, i.e. $T_\text{r}\geq T_\text{a}$.}

With the definition of semi-MDP, it is possible to separate the original problem into sub-problems: scheduling and association. They can be jointly optimized by a multi-layer hierarchy structure with scheduling as meta-controller \cite{Kulkarni}. First, we write our optimization target as maximizing the accumulative V-PSNR gain over broadcast slots in $T_{\text{GOP}}$ via finding the optimal $\pi_\text{s}$ and $\pi_\text{a}$.
\begin{equation}
    \max\limits_{\pi_{\text{s}}, \pi_{\text{a}}} \mathds{E}{[\sum^{T_{\text{GOP}}}_{t_{\text{b}}=0}\underbrace{\sum_{j\in J_{t_\text{b}}}\underbrace{\sum_{v\in\mathcal{V}_j}\Delta \text{V-PSNR}^v_{t_{\text{b}}}}_{\text{V-PSNR Gain for transmitted tile } j}}_{\text{V-PSNR Gain in } T_\text{b}\text{ for scheduled tile set } J_{t_\text{b}}} | J_{t_\text{b}}\sim \pi_\text{s}, \mathcal{V}_j\sim\pi_\text{s}]},
    \label{optimize}
\end{equation}
where the V-PSNR gain is denoted as $\Delta \text{V-PSNR}^v_{t_\text{b}} = \text{V-PSNR}^v_{t_\text{b}} - \text{V-PSNR}^v_{t_\text{b}-1}$. 
The scheduling sub-problem acts as a meta-controller to optimize the cumulative intrinsic V-PSNR gain with certain $\pi_{\text{a}}$ in $T_\text{r}$ time-scale:
\begin{equation}
    \max\limits_{\pi_{\text{s}}} \mathds{E}{[\sum^{T_{\text{GOP}}}_{t_{\text{r}}=0}\underbrace{\sum_{j\in J_{t_\text{r}}}\sum_{v\in\mathcal{V}_j}\Delta \text{V-PSNR}^v_{t_{\text{r}}}}_{\text{V-PSNR gain within } {T_\text{r}}} | \pi_{\text{a}}]}.
    \label{optimize_sche}
\end{equation}
The association sub-problem maximizes the cumulative extrinsic V-PSNR gain with a given $J_{t_\text{b}}$ in $T_\text{b}$ time-scale
\begin{equation}
    \max\limits_{\pi_{\text{a}}} \mathds{E}{[\sum^{T_{\text{GOP}}}_{t_\text{b}=0}\underbrace{\sum_{j\in J_{t}}\sum_{v\in\mathcal{V}_j}\Delta \text{V-PSNR}^v_{t_{\text{b}}}}_{\text{V-PSNR gain within } T_\text{b}}]}.
    \label{optimize_asso}
\end{equation}
From \eqref{optimize_sche} and \eqref{optimize_asso}, we can observe that the scheduling and association problems are directly coupled, which need to be jointly optimized.

\subsection{Conventional Approaches}
In this section, we introduce conventional scheduling and association approaches for each sub-problem, namely, popularity-based proportional fair (P-PF) scheduling, cell-based (CB), and cell-free (CF) associations, respectively.

\subsubsection{Popularity-based Scheduling}
According to \eqref{decode_tile} and \eqref{optimize_sche}, the potential V-PSNR gain for transmitting the tile $j$ is jointly determined by the number of VR users in the group $\mathcal{V}_j$, and the transmission successful rate in current and previous broadcast slots. From \eqref{totalTarget_rewrite} and \eqref{decode_tile}, we know that the V-PSNR gain in each broadcast slot $T_\text{b}$ tightly correlates to the number of VR users who request the tile $j$, i.e. $|\mathcal{V}_j|$. Thus, the more VR users request the tile $j$, the more V-PSNR gain via transmitting the tile $j$. This instantly results in a popularity-based scheduling algorithm, where tiles with higher popularity are transmitted in each $T_\text{b}$. Remind that, the scheduling action directly decides which tile to transmit in each broadcast slot for each UAV, which in turn decides and the corresponding VR user group $\mathcal{V}_j$.

Additionally, to take decoding state and fairly serve all VR users, we borrow the idea of proportional fair (PF) scheduler that has been widely used in existing cellular network~\cite{Margolies2016}. By adding the previous tiles' decoding state in denominator, the resulting P-PF scheduling method determines the prioritization of tile $j$ at time ${t_\text{r}}$ as 
\begin{equation}
    \text{P-PF}_{j} = 
    \frac{\sum\limits_{v\in\mathcal{V}_j} \mathds{1}[j \in \mathcal{J}^{v}_{{t_\text{r}}},j\notin\mathbf{J}^v_{t_\text{r}}]}{\sum\limits^{{t_\text{r}}-1}_{{t_\text{r}}'=0} \mathds{1}[j_{t_\text{r}'} \in \mathbf{J}^{v}_{t_\text{r}}]},
\end{equation}
where $\mathbf{J}_{t_\text{r}}^v$ denotes the successfully decoded tiles in the $v$th VR user at time $t$, and $j_{t_\text{r}'}$ denotes the tile at time $t_\text{r}'$ that is required by tile $j$'s decoding, the value of numerator is 1 if current tile is required by $v$th VR user, the value of denominator is the sum of previous successfully received tiles, which is required by $j$ to decode.

\subsubsection{Cell-based and Cell-free Association}
We adopt two conventional network schemes to handle the association problem, which are cell-based (CB) and cell-free MIMO (CF) associations: 1) In CB network, each AP is an individual cell, where each AP makes its decision based on its observation independently cooperation. Specifically, each AP is associated with the largest VR user group $\mathcal{V}_j$, which has its corresponding $v$th UAV and $j$th tile ($j\in\mathcal{J}^v$) inside its observation. This scheme may bring high inter-cell interference and poor cell-edge performance; and 2) In the CF network, all APs cooperatively receive one tile in every \emph{UAV-APs} uplink stage and broadcast one tile in the broadcast stage. In another word, all APs are grouped in one virtual cell. In this scheme, the tile with the highest priority among all tiles in all UAVs is selected to be transmitted. This scheme provides high channel capacity for transmitting UAV and corresponding VR users, resulting in inefficient time resource usage with geometry correlated VR users, i.e. $D_{u,v}\gg\mu M_\text{T}$.

\section{Reinforcement Learning Approach for Association}
With separated sequential scheduling and association sub-problems, we first design an intelligent association algorithm working with the conventional P-PF scheduling method to showcase the benefits of adjusting the association dynamically. By employing the deterministic scheduling algorithm, the original problem now degrades as a common MDP. This eases our analysis. 

In our considered scenario, the geometry-correlated VR users' requests provide another degree-of-freedom in system design. \textcolor{black}{The association algorithm should spatially reuse the frequency resource by dynamically grouping APs into virtual cells, which improves the resource utilization and efficiency of the system. The conventional approaches are simple and easy to deploy, but their performances drop in certain scenarios due to the lack of adjustment based on the environment. It calls for an intelligent algorithm, which is capable of adjusting association policy for the complex and high-dimensional environment with hundreds of VR users. Among different intelligent algorithms, reinforcement learning is shown to be useful in solving communication problems, which are model-free and shown to be useful in addressing POMDP problems with a complex environment \cite{Peng2021Multi,Shiri2020Comm}.}

\textcolor{black}{
To solve the association problem with reinforcement learning algorithms, we notice that the state information can't be fully observed. Both the future channel state information and the precise information of the users' positions are unavailable. In this way, the problem has to be updated as a partially observable MDP (POMDP). We then complete the definition of our considered POMDP with the following definitions:}:
\begin{itemize}
    \item The observation $o$ ($o\in\mathcal{O}$) only contains all nodes' position and VR users' tile request without including the UAV and VR users' channel state, due to that the channels are largely dominated by the large-scale fading under APs' cooperative reception and transmission in CF-MB network \cite{Ngo2017}.
    \item The action $a$ ($a\in\mathcal{A}$) for the association is a one-hop mapping from each AP to the tuple of serving UAV, tile $j$, and corresponding VR user group $\mathcal{V}_j$. As each AP has $|\mathcal{U}|$ actions to choose, the size of action space of $\mathcal{A}$ can be calculated as $|\mathcal{A}| = |\mathcal{U}|^{|\mathcal{B}|}$, and the action at time $t$ is denoted as $A_t$. 
    \item The reward $R_t$ ($R_t=r, r\in\mathcal{R}$) is the V-PSNR gain at time $t$, designed as
    \begin{equation}
        R_{t}= \sum_{v\in\mathcal{V}}\Delta\text{V-PSNR}_{t}^v,
        \label{reward_centralized}
    \end{equation}
    where $\Delta\text{V-PSNR}_{t}^v = \text{V-PSNR}_{t}^v-\text{V-PSNR}_{t-1}^v$.
\end{itemize}

\begin{figure}[t]
    \begin{minipage}[t]{0.48\textwidth}
    \centering
    \includegraphics[width = \textwidth]{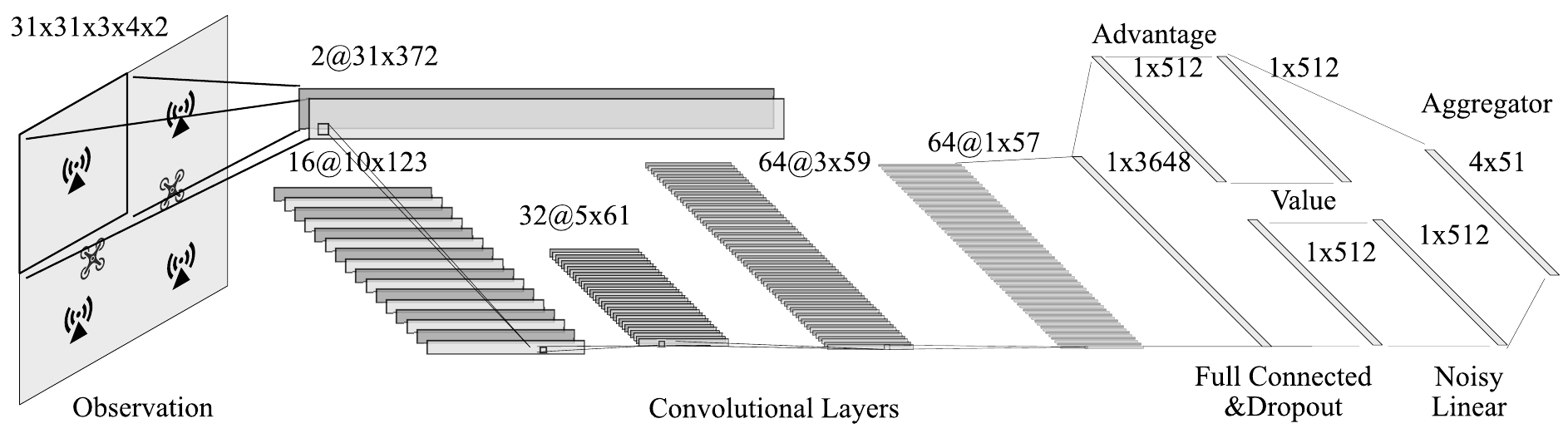}
    \caption{Network Structure of Distributed Association Agent.}
    \label{fig:asso_learning}
    \end{minipage}
    \hspace*{0.4cm}
    \begin{minipage}[t]{0.48\textwidth}
    \centering
    \includegraphics[width = \textwidth]{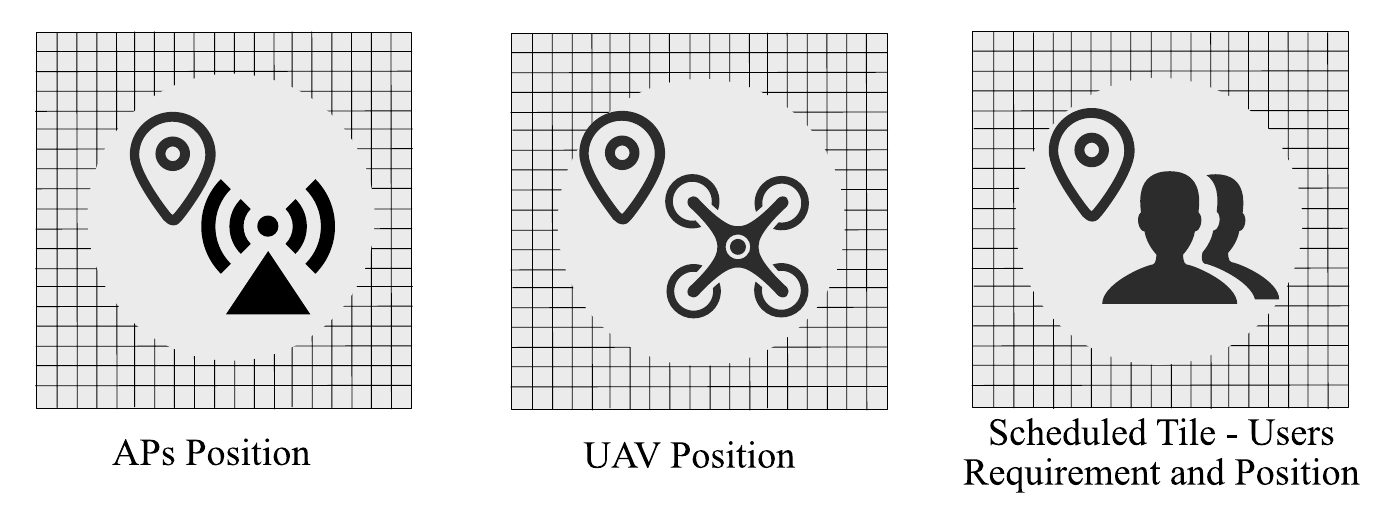}
    \caption{The grid-based observation generated from environment state.}
    \label{fig:observationl}
    \end{minipage}
\end{figure}

\subsection{Centralized Deep Reinforcement Learning}
To solve our proposed POMDP problem with a reinforcement learning algorithm, we start from a very basic centralized approach to proving the effectiveness and set a baseline for the learning approaches. Remind that in our considered CF-MB network, the existence of the central server naturally facilities the centralized approach, where a centralized agent is placed at the central server to make joint association decisions for all APs dynamically and maximize the long-term V-PSNR. \textcolor{black}{Considering that the state space is too large and impossible to be captured via a conventional table-based reinforcement learning approach. The deep neural networks are introduced, which effectively encode and represent the observation as a low dimensional hidden vector by discovering the similarity between them. With a small size hidden vector, the following neural layers can effectively fit the target value with small network size.}

\subsubsection{Grid-based Observation and Neural Network Layers}
To further reduce the complexity of observation, we manually degraded the input dimension via a grid-based observation, which is a fuzzy representation of the state. For each pixel-like grid, the geometry-correlated observed information inside is summed and presented. As shown in Fig. \ref{fig:observationl}, for each UAV, we have three grid-maps, which correspond to the position of UAVs, APs, and the VR user group requesting the currently scheduled tiles ($\mathcal{V}_j, j\in J_t$), respectively. The value in each UAV and APs grid map is $1$ if the node exists in that grid. For the VR user group grid-map, the value in each grid is the summation number of tile requests from the VR users in that grid, which is normalized into the range of $(0,1]$ over the maximum number of tiles' requests in each grid. For example, two VR users from $1$th UAV locate in the same grid and request $2$ and $3$ tiles from scheduled tiles $J_t$, the maximum number of tile requests in grids is $8$. Then, the normalized value in that grid is $0.625$.

To capture the spatial information in grid observations among UAV, AP, and VR users, we introduce convolutional layers to encode the observation into a low dimensional vector. The benefit of applying convolutional layers in communication problems has been shown by previous researches \cite{Cui2019}. The convolutional layers can easily learn to estimate the potential signal and interference, as the convolutional operation matches the signal and interference calculation formula. As shown in Fig. \ref{fig:asso_learning}, we design five layers of convolutional layers and one linear layer to encode the observation into a hidden vector, which is then processed by a duelling network. The duelling network contains two streams composed of two noisy linear layers: advantage and value stream, respectively. The advantage stream measures how good the action will be compared to the averaged V-PSNR. The value stream gives the expectation of V-PSNR value from the current state. The output from both streams is then aggregated as the policy \cite{Hessel2018}.

\subsection{Rainbow Algorithm}
\textcolor{black}{However, the conventional neural network in reinforcement learning is designed for static reward from the environment \cite{jaakkola1995reinforcement}. Due to the random nature of the wireless environment and the absence of a channel state, the reward varies in distribution form. The distributional DRL approach allows the algorithm to adapt to this case, which improves the performance \cite{Hessel2018}.
With distributional DRL, the value function's distribution, which is denoted as $z(s,o,a)$, is directly estimated in distribution form. The value estimation $d$ generated by the neural network for each action is a discrete mapping from the actual value distribution $z(s,o,a)$ to $N_\text{atom}$ distributive value supports.} We denote the distribution mapping as $d = (z, p_\theta(s, o, a))$, with probability mass $p^i_\theta(s, o, a)$ on $i$th support. Then, the action with the highest expectation of the estimated distribution is selected. The network parameters $\theta$ are updated and optimized by minimizing the Kullbeck-Leibler divergence between the estimated distribution (estimated by neural network with parameter $\theta$) and target distribution $d_t$ at time $t$ as \cite{Bellemare2017}
\begin{equation}
\begin{aligned}
    D_{\text{KL}}(\sum_{k=0}^{n-1} R_{t+k+1} + z, p_\theta(S_{t+n}, O_{t+n}, a') || d_t),
    \label{learning_loss}
\end{aligned}
\end{equation}
which measures the difference between forward-view $n$-step distribution target and current distribution estimation $d_t$ at time $t$, the $a'$ ($a'\in\mathcal{A}(S_{t+n})$) is the action selected by the policy and estimated distribution from neural network with parameter $\theta$ and at time $t+n$. The loss is minimized with categorical algorithm and gradient descent \cite[Algorithm. 1]{46981}. We select the algorithm which provides the distributional estimation capability as well as other stability improvement tricks, such as double Q-learning, which is called rainbow \textbf{Algorithm. \ref{algo_rainbow_association}}.

\begin{algorithm}[t]
\setstretch{0.9}
\DontPrintSemicolon
\SetKwData{Left}{left}\SetKwData{This}{this}\SetKwData{Up}{up}\SetKwFunction{Vpsnr}{V-PSNR}\SetKwFunction{Gamend}{Game end}\SetKwFunction{Reset}{Reset}\SetKwFunction{Rot}{Rotation}\SetKwFunction{Plt}{Plot}\SetKwFunction{Sche}{Scheduling}\SetKwFunction{Eval}{Evaluation}\SetKwInOut{Input}{input}\SetKwInOut{Output}{output}

\Input{An environment $Env$.}
\BlankLine
Initiate network parameters.\;
Initiate environment $Env$, state $S_0$ and observation $O_0$.\; 
\Repeat{Converge}{
\If{\Gamend}{
Obtain $S_0$ from revising environment \Reset{$Env$} and set $t=0$\;
}
\If{$t$ can be divided by $T_\text{r}/T_\text{b}$}{
Obtain network observation $O_t$ and scheduled tile set $J_t$ for current time period from \Sche{$S_t$}\;
}
Select an action $A_t$ greedily: $A_t=\argmax_{a\in\mathcal{A}(S_t)} \mathds{E}[d_t]$\;
APs forms virtual cells $\mathcal{B}^u_t$ based on action $A_t$\;
Tile is transmitted from $u$th UAV to corresponding VR users set $\mathcal{V}^u$ via APs group $\mathcal{B}^u_t$. \;
$Env$ generates new state $S_{t+1}$\;
Calculate reward $R_t$ for all VR users\;
Push tuple $(O_t,A_t,R_t)$ to experience replay\;
Steps time period index $t\leftarrow t+1$\;
Train the network parameters by minimising loss defined in \eqref{learning_loss} with a batch of memories $(O_{t'},A_{t'},R_{t'},O_{t'+1})$ in experience replay\;
Perform a gradient descent for neural network\;
}
  \caption{Rainbow DRL based APs association.}
  \label{algo_rainbow_association}
\end{algorithm}

\subsection{Networked-Markov Decision Process and Distributed Multi-Agent Algorithm}
It is important to note that the performance of a centralized learning approach is largely limited by the dimension explosion problem caused by increasing serving area and the number of participating APs, i.e. the action space grows exponentially with the number of APs $|\mathcal{A}| = |\mathcal{U}|^{|\mathcal{B}|}$. Besides, the transmission, concatenation, and processing of large size observation at the central server cause heavy backhaul overhead.

To address this issue, we apply distributed reinforcement learning, where each agent makes its own decision individually. We then further divide the association optimization target Eq. \eqref{optimize_asso} spatially and solves it via a homogeneous multi-agent setting based on the mean-field theorem \cite{Yang2018,shiri2020communication}. \textcolor{black}{First, the wireless signal fades with the increase of communication distance, especially for our considered small APs. The far-side UAVs and APs have limited impacts on the signal gain or interference of the current AP's surrounding area. Second, the reward function is geometrically separable, and each AP can obtain a precise part that is correlated to it in global reward. Third, each AP always has a limited amount of neighbours, which is far smaller than the overall number of agents. Third, the number of correlated AP is small compared to the overall AP number. Thus, as shown in Fig. 3, it is possible to let each agent only capture the observation from surrounding areas without losing any useful information, since the surrounding area contains all information that correlates to the current AP. The surrounding areas of different APs are partly overlapped (cell-edge area). As such, the set of AP forms a network and each AP only cares about itself and its neighbours.}

With the above characteristics, we can formulate our association problem as a networked decentralized partially observable Markov decision processes (ND-POMDP) problem \cite{Ranjit2005}, which is a factored version of Decentralized-POMDP problem with mean-field theorem~\cite{Yang2018}.
The local observation set $\mathcal{O}_b$ now contains the local observation information surrounding $b$-th agent. The joint action space can be denoted as $\mathcal{A}=\prod_{b\in\mathcal{B}^b}\mathcal{A}_b$, where $\mathcal{A}_b$ is the set of local action space of the $b$th AP. The reward for $b$th AP is denoted as $R_t^b(s,a_b,\mathbf{a}_{-b})=\sum_{v\in\mathcal{V}^b}\Delta \text{V-PSNR}_t^v$, where $\mathcal{V}^b$ is the VR users in $b$th AP's observation range. 

Then, the Bellman equation for $b$-th agent with state-action function $q_b(s,a_b)$ can be written as:
\begin{equation}
\begin{aligned}
    q_b(s,a_b)&= \sum_{j\in J_t}\sum_{v\in\mathcal{V}_j}\Delta \text{V-PSNR}_t^v \\ &+\mathds{E}_{s'\in\mathcal{S}}[\sum_{a_{b}'\in\mathcal{A}_b}\pi_\text{a}(a_b'|s',(\mathbf{a}_{-b}))q_b(s',a_{b}')].
\end{aligned}
\label{distribute_value}
\end{equation}
where the $\pi_{a}^{-b}$ present the joint policy of $b$th agent's neighbors, $s'$ is the state at $t+1$, $\mathbf{a}_{-b}$ presents $b$th agent's neighbors' action. As such, the size of the problem is largely reduced and can be solved distributively. Each agent improves the V-PSNR from surrounded VR users which also improve the overall V-PSNR value. \textcolor{black}{The DN-POMDP can be solved by common reinforcement learning approaches and proved to converge \cite{lowe2017}, as the environment is stationary with known neighbours' policy. The algorithm is also shown to be converged with averaged neighbours policy \cite{Yang2018}.}

\textcolor{black}{However, from Eq. \eqref{distribute_value}, we can see that the optimization target still depends on $b$th AP's neighbours' policy, which is not controllable for the current AP. In some cases, sharing actions among agents is undesirable due to latency or privacy reasons. In our case, we consider the agent without the neighbours' information. The environment is a non-stationary environment from the perspective of any individual AP. Here, we employ federated learning and Boltzmann policy to reduce the variance of the learning. 
Federated learning has been shown useful in improving cooperative performance~\cite{Ranjit2005}. In our considered network, the optimization problem for all agents can be seen as identical with a similar environment and reward. Federated learning can improve the learning speed and reduce the variance caused by unknown neighbours' policy \cite{8792117}. Although the convergence of federated learning multi-agent algorithm with neighbours' actions is similar to parameter sharing of multi-agent learning and shown by many works \cite{9174775}, but there is no strict proof of that without neighbours' actions and only shown to be useful in practice.
We apply FL via federated average (FedAvg) algorithm which performs averaging every $T_\text{federated}$ time intervals. }
The second trick we used is Boltzmann policy. Greedy action selection is widely used in reinforcement learning algorithm. However, in multi-agent setting, the action with maximum value usually requires other agents' cooperation. This usually does not hold while all agents selecting their action greedily \cite{Yang2018}. Thus, the greedy action selection ignores the need of potential cooperation actions from neighbors, which can easily fail to converge. Thus, we adopt the Boltzmann policy to capture actions with relatively small return, but potentially benefit the overall environment via effective cooperation. The Boltzmann policy for $b$th AP in state $s$ can be formulated as \cite{Yang2018}
\begin{equation}
    \pi_{\text{a}}^b(a_b|s, (\mathbf{a}_{-b})) = \frac{\exp{(-\beta q_b(s, a_b))}}{\big(\sum_{a_b\in \mathcal{A}_b}\exp{(-\beta q_b(s, a_b))}\big)},
    \label{boltz_action}
\end{equation}
where $\beta$ is the temperature for Boltzmann policy, $q_b(s, a_b)$ is the estimation output of the network at $a_b$ action. To solve the optimization problem for each separated sub-problem, we again adopt a rainbow agent for the same reason as the centralized learning approach. The algorithm is presented in \textbf{Algorithm. \ref{algo_rainbow_hier}}. 

\begin{algorithm}
\setstretch{0.9}
\DontPrintSemicolon
\SetKwData{Left}{left}\SetKwData{This}{this}\SetKwData{Up}{up}\SetKwFunction{Vpsnr}{V-PSNR}\SetKwFunction{Gamend}{Game end}\SetKwFunction{Reset}{Reset}\SetKwFunction{Rot}{Rotation}\SetKwFunction{FedAvg}{FedAvg}\SetKwFunction{PO}{Partially Observe}\SetKwFunction{Eval}{Evaluation}\SetKwInOut{Input}{input}\SetKwInOut{Output}{output}

\Input{An environment $Env$}
\BlankLine
Initiate \emph{scheduling network} and \emph{association network} parameters.\;
Initiate environment $Env$, state $S_0$, and scheduling observation $O_0^\text{s}$.\; 
\Repeat{Converge}{
\If{\Gamend}{
Reset $Env$ and $t=0$, obtain new $S_0$, $O_0^\text{s}$\;
}
\If{$t$ can be divided by $T_\text{r}/T_\text{b}$}{
Store tuple $(O^\text{s}_{t-N}, A^\text{s}_{t-N}, \sum_{t'=t-N}^t R_{t'})$ \KwTo scheduling experience replay\;
Calculate priority of scheduling $A^\text{s}_t=\mathds{E}[d^\text{s}_t]$\;
Select ${T_\text{r}/T_\text{b}}$ tiles with largest priority for each UAV $J_t = \argmax (A^\text{s}_t, {T_\text{r}/T_\text{b}})$\;
Train and update \emph{scheduling network}'s parameters with memories in experience replay\;
}
\For{$b\in\mathcal{B}$}{
Obtain $O^b_t$ from state $S_t$, scheduled tiles $J_t$ and past state $S_{t-1}$\;
Select an action $A^b_t$ with \eqref{boltz_action}\;
}
APs forms virtual cells $\mathcal{B}^u_t$ for UAVs based on their actions\;
Tiles are transmitted from $u$th UAV to corresponding VR users $\mathcal{V}^u$ via APs group $\mathcal{B}^u_t$\;
\For{$b\in\mathcal{B}$}{
Calculate reward $R^b_t$ for VR users in $b$th AP's surrounding area\;
Push tuple $(O^b_t,A^b_t,R^b_t)$ to $b$th AP's experience replay\;
Train and update \emph{association network}'s parameters following the same procedure as Algorithm. \ref{algo_rainbow_association}.\;
}
Step $Env$ and generates $S_{t+1}$.\;
\If{$t$ can be divided by $T_\text{federated}$}{
Perform \FedAvg among APs $\mathcal{B}$\;
}
}
  \caption{Hierarchical DRL based joint scheduling and association.}
  \label{algo_rainbow_hier}
\end{algorithm}

\section{Hierarchical Learning with Learning-Based Scheduler}
After solving the association problem, we try to include scheduling in an intelligent algorithm and investigate the benefit of the joint design of the scheduling and association. Recap that the scheduling and association process is executed successively, we define the joint problem as a semi-ND-POMDP with scheduler as options of the association process's ND-POMDP. We then try to solve the problem with a hierarchical reinforcement learning algorithm.

\begin{figure*}
    \centering
    \includegraphics[width=0.9\textwidth]{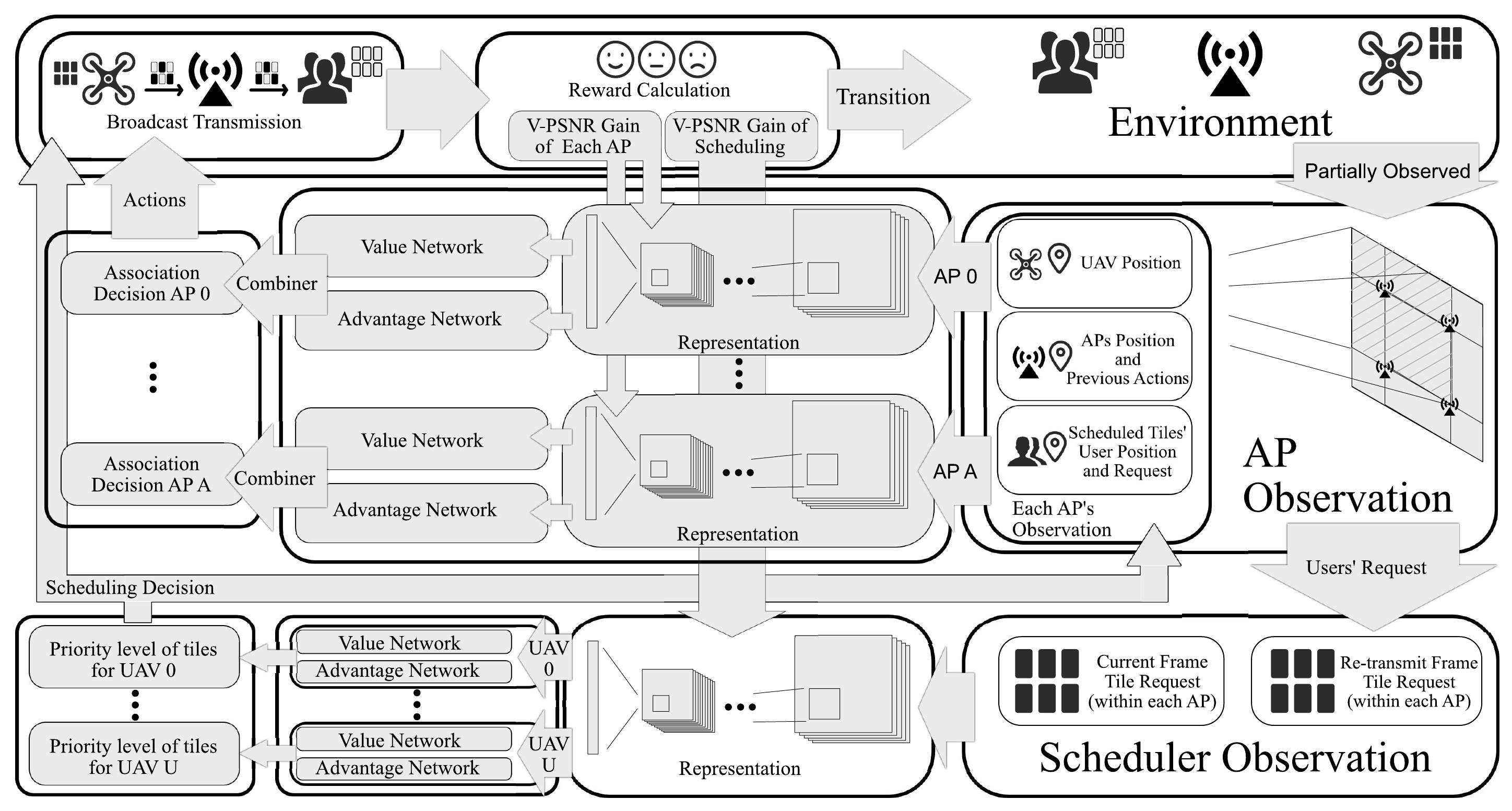}
    \caption{Learning Architecture of Hierarchical Reinforcement Learning}
    \label{fig:hac_struc}
\end{figure*}

The idea of hierarchical reinforcement learning is updating two networks, one for scheduling policy and another for association policy. We update the scheduling policy with a fixed association policy, verse visa. Note that both scheduling and association network shares the same environment reward in different time scale. This guarantees monotonic improvement for the environment V-PSNR. In the transmission procedures, each AP firstly observe the tile requests and UAVs', APs', and VR users' positions from the environment. The agent of the scheduler at a central server then generates its observation based on the returned information from APs. The $J_t$ tiles are then scheduled for transmission during the next re-scheduling slot. The agents constantly adjust their association decisions based on their policy, and the AP capture the V-PSNR gain to improve the scheduling and association policy. Recap that the scheduling option part contains initial state, policy and scheduling priority action set which is denoted by $\mathcal{I}, \pi^s, \mathcal{A}^\text{s}$, respectively.
As such, following the optimization target in Eq. \eqref{optimize_sche}, the Bellmann equation for scheduling part with option-value function $q_{\text{s}}(s,a_{\text{s}})$ can be written as:
\begin{equation}
    \begin{aligned}
    q_{\text{s}}&(s,a_{\text{s}})=\sum_{k=t}^{t+N}\sum_{j\in J_k}\sum_{v\in\mathcal{V}_j}\Delta \text{V-PSNR}_k^v \\
    &+\mathds{E}_{s'\in\mathcal{S}}[\sum_{a_{\text{s}}'\in\mathcal{A}_s}\pi_\text{s}({a_{\text{s}}'}|s)q_{\text{s}}(s',a_{\text{s}}') | J_k\sim\argmax a_\text{s} s'\in\mathcal{I}],
    \end{aligned}
\end{equation}
where option-value function $q_{\text{s}}(s,a_{\text{s}})$ presents the expected future reward, $N$ is the number of system time period before re-scheduling, $a_\text{s}$ is the scheduling priority and the action of scheduling part. Similar to the algorithm in Section IV, the Bellman equation of $b$th AP's association with scheduled tiles $J_t$ at time $t$ can be rewritten as
\begin{equation}
    \begin{aligned}
    q_b(s,a_b;J_t)&= \sum_{j\in J_t}\sum_{v\in\mathcal{V}_j}\Delta \text{V-PSNR}_t^v \\ &+\mathds{E}_{s'\in\mathcal{S}}[\sum_{a_{b}'\in\mathcal{A}_b}\pi_\text{a}(a_b'|s',(\mathbf{a}_{-b}))q_b(s',a_{b}';J_{t+1})].
    \end{aligned}
\end{equation}
The association policy maximizes the intrinsic reward, which is the V-PSNR gain in each broadcast slot. The scheduling policy maximizes the extrinsic reward, which is the potential V-PSNR gain for transmitting scheduled tiles with a certain association policy.

\begin{figure}
    \centering
    \includegraphics[width = 0.5 \textwidth]{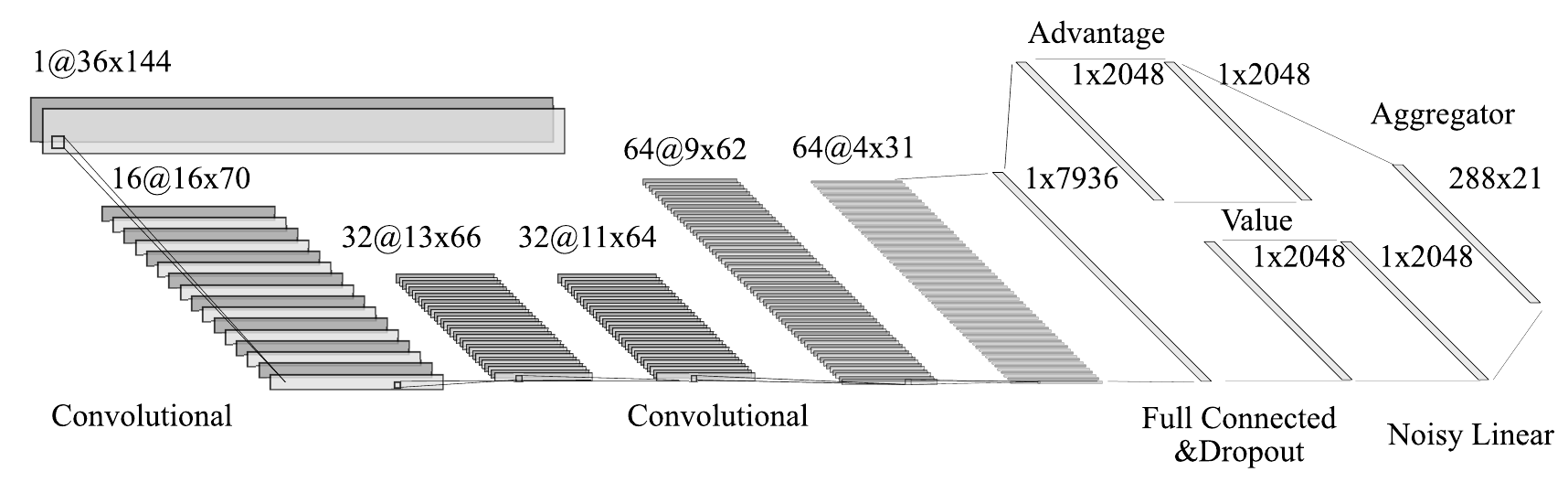}
    \caption{Network Structure of Scheduler Agent in Center Server}
    \label{fig:sche_learning}
\end{figure}

To capture the complex environment for scheduling and association, as shown in Fig. \ref{fig:hac_struc}, we follow the same association optimization algorithm as the distributed DRL approach in Section IV. For the scheduling part, we also employ the CNN-based rainbow algorithm with a smaller reception field than the association part for the same reason. The tile requests are naturally clustered as VR users always request continue tiles inside their FoV. Scheduling the nearby tiles can potentially benefit the V-PSNR by fully completing the tile request from part of VR users first. Due to the random nature of the communication system, the V-PSNR gain of transmitting groups of tiles is in distribution form.

With hundreds of tiles and corresponding requests, the observation of the scheduling part is composited by the popularity of tiles in $6\times 12$ grid where the popularity locates in the same position as they are in $360\degree$ view. The popularity of VR users in different locations is then concatenated into a joint popularity map. Apart from popularity, we also consider the popularity of re-transmission tiles in the same formula. Thus, each AP observes $6 \times 12 \times 2$ tiles' popularity for each UAV from its surrounding state. The overall observation is generated by concatenating each AP's observation. 
Thus, taking one example, the serving area is separated into $3\times 3$ squares. Thus, there are $3\times 12$ grid in the horizontal axis and $3\times 6 \times 2$ in the vertical axis for each UAV's tiles' request. Then, as shown in Fig. \ref{fig:sche_learning}, we apply a similar network structure as the agent in Section IV. The $|\mathds{J}_t|$ tiles with the highest weight are scheduled and transmitted. The network makes an association decision and transmits tiles. Then, the network is updated with returned V-PSNR gain. The algorithm for the hierarchical learning approach is represented in \textbf{Algorithm. \ref{algo_rainbow_hier}}.

\section{Simulation Result}
In this section, we examine the QoE of tile streaming from UAV to VR users in our proposed CF-MB network within a squared serving area.
The parameter of our simulation and learning system is given in Table. \ref{tab:paramters}\footnote{The authors acknowledge the use of the research computing facility at King’s College London, Rosalind (\url{https://rosalind.kcl.ac.uk}).\label{rosalind}}. In the following, we present the V-PSNR performance for our proposed three learning algorithms in Section VI-A and Section VI-B.


In the simulation, we set the number of VR users as $|\mathcal{V}|=120$, the VR users are distributed following PCP, whose cluster radius is set as $r_\text{c}=20m$, the number of UAVs is $|\mathcal{U}|=4$. We set the number of AP as $|\mathcal{B}|=9$, which are located in a $3\times3$ grid with \SI{30}{\meter} gap inside the serving area which is \SI{80}{\meter}$\times$\SI{80}{\meter} square. Each AP can observe $60m\times 60m$ squared area surrounding itself. The time period of learning algorithms contains $10T_\text{b}$, which means the scheduling and association policy is updated after broadcasting $10$ tiles. \textcolor{black}{Note that for a centralized algorithm, due to the large action space of our environment setting ($|\mathcal{A}|=4^9$), we can't train this oversize model with commercial computers. Thus, we reduce the size of action space by only 2 UAV and half broadcast slots of the current setting. We plot the performance of a centralized algorithm just to show the effectiveness of centralized learning in this scenario. Note that it does not present the actual performance of centralized learning with the full-size environment.}

\begin{table*}[t]
\centering
\resizebox{1.85\columnwidth}{!}{%
\begin{tabular}{|l|l|l|l|}
\hline
\rowcolor{LightGray}
\textbf{Channel parameters} & \textbf{Setting} & \textbf{Channel parameters} & \textbf{Setting} \\ \hline
AP-VR link path-loss exponent (AP-VR) $\alpha_{\text{DL}}$ & 4 & UAV-AP link path-loss exponent (UAV-AP) $\alpha_{\text{UL}}$ & 2 \\
VR center frequency & \SI{5.5}{\giga\hertz} & UAV center frequency & \SI{4.5}{\giga\hertz}\\
Accesspoint grid length & \SI{30}{\meter} & Drone hovering height & \SI{30}{\meter}\\
User density & 100 & Excessive NLoS Attenuation & \SI{20}{dB}\\
Accesspoint EIRP & \SI{48}{dBm} & UAV EIRP & \SI{48}{dBm} \\
Accesspoint transmission bandwidth & \SI{5}{\mega\hertz} & UAV transmission bandwidth & \SI{5}{\mega\hertz} \\
Noise power $\theta^2$ & \SI{-91}{dBm} & Number of UAV & 2 \\
\hline
\rowcolor{LightGray}
\textbf{Video parameter} & \textbf{Setting} & \textbf{Learning parameters} & \textbf{Setting} \\ \hline
Frame rate & \SI{90}{\hertz} & Temperature ($\beta$) & 100 \\
Group of picture & 5 (IPPPP) & Learning rate & $6.25\times 10^{-5}$ \\
Pixel per degree & 60 & Dropout rate & 0.2 \\
Video compression rate & 150 & Batch size & 32 \\
Frame size ratio (P/I) & 0.7 & Atoms (Association) & 21 \\
User field-of-view & $210\degree\times 150\degree$ & Atoms (Scheduler) & 11 \\
Tile size & $30\degree\times 30\degree$ & Noisy layer std & 0.5 \\
Number of re-schedule between frames $T_{\text{f}}$ & 28 & Discount $\gamma$ & 1 \\
Broadcast slots between re-schedule $T_\text{r}$ & 10 & Multi-step learning & 3 \\
 \hline
\end{tabular}
}
\caption{Environment and Learning Parameters.}
\label{tab:paramters}
\end{table*}


\begin{figure}[t]
    \begin{minipage}[t]{0.48\textwidth}
    \centering
     \begin{subfigure}[b]{0.48\textwidth}
         \centering
         \includegraphics[width=\textwidth]{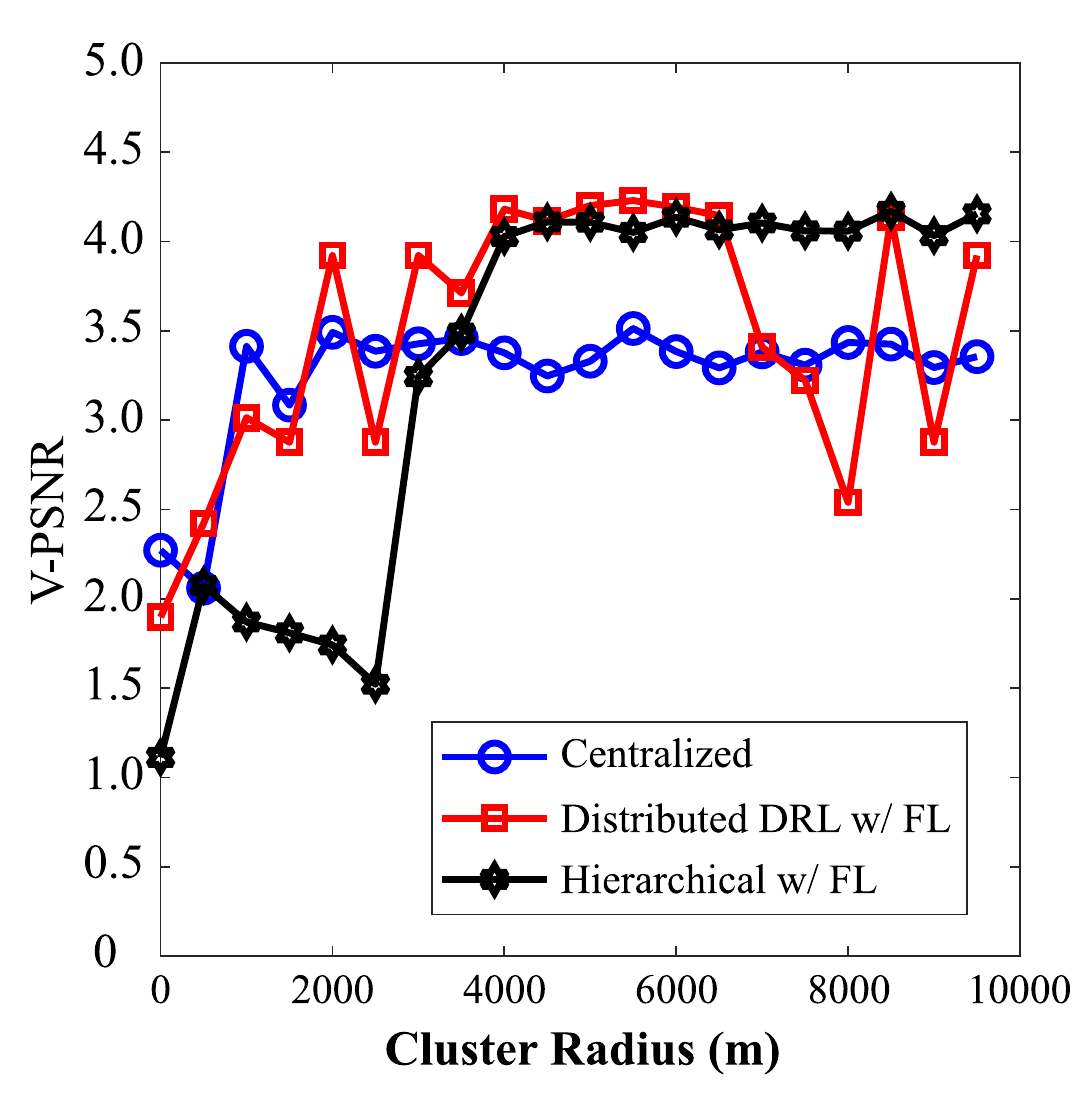}
         \caption{}
         \label{fig:psnr_3}
     \end{subfigure}
     \hfill
     \begin{subfigure}[b]{0.48\textwidth}
         \centering
         \includegraphics[width=\textwidth]{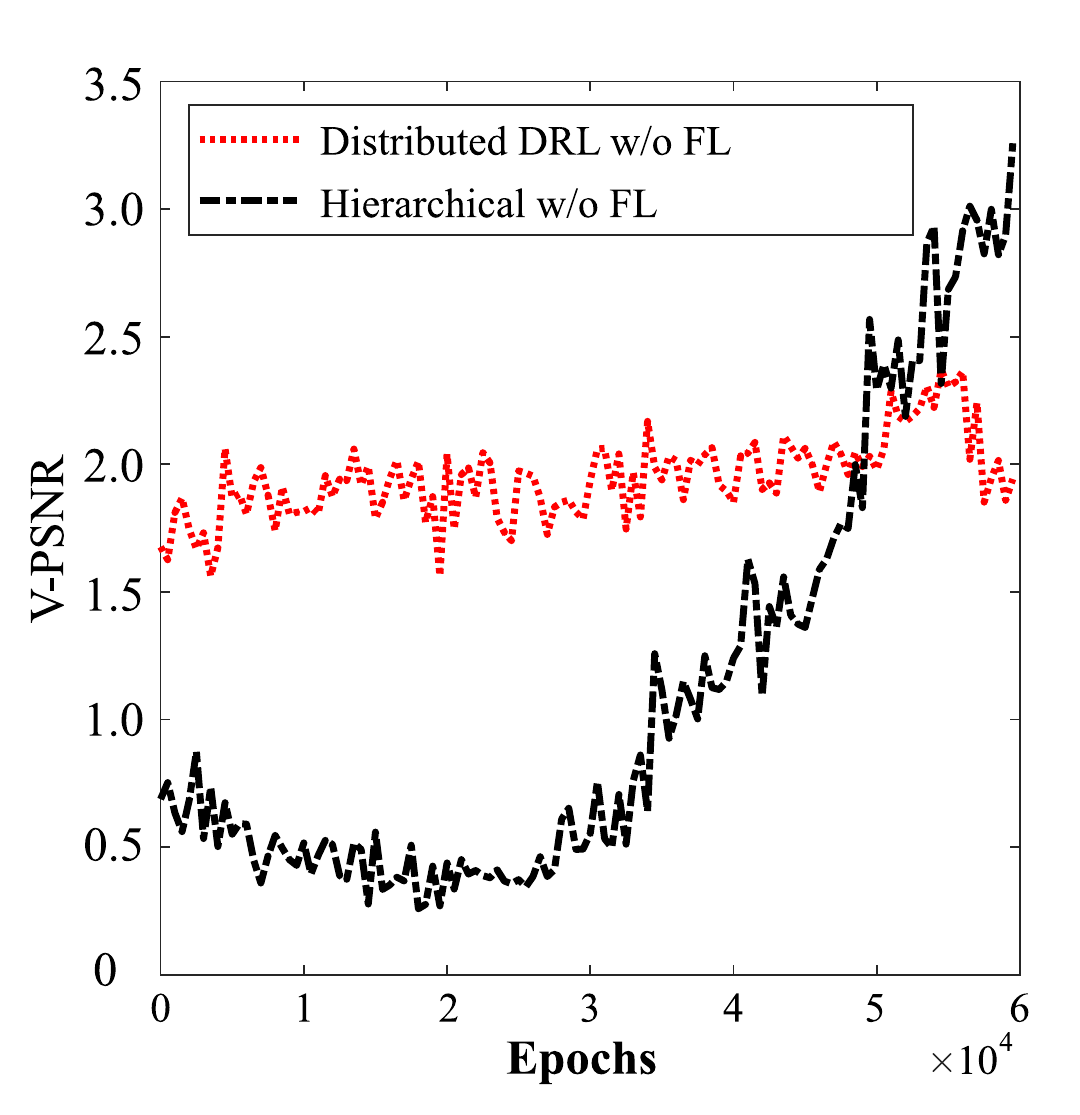}
         \caption{}
         \label{fig:psnr_33}
     \end{subfigure}
     \caption{Convergence curves for our proposed algorithms.}
     \label{fig:psnr}
    \end{minipage}
    \hspace*{0.4cm}
    \begin{minipage}[t]{0.48\textwidth}
    \centering
    \includegraphics[width = \textwidth]{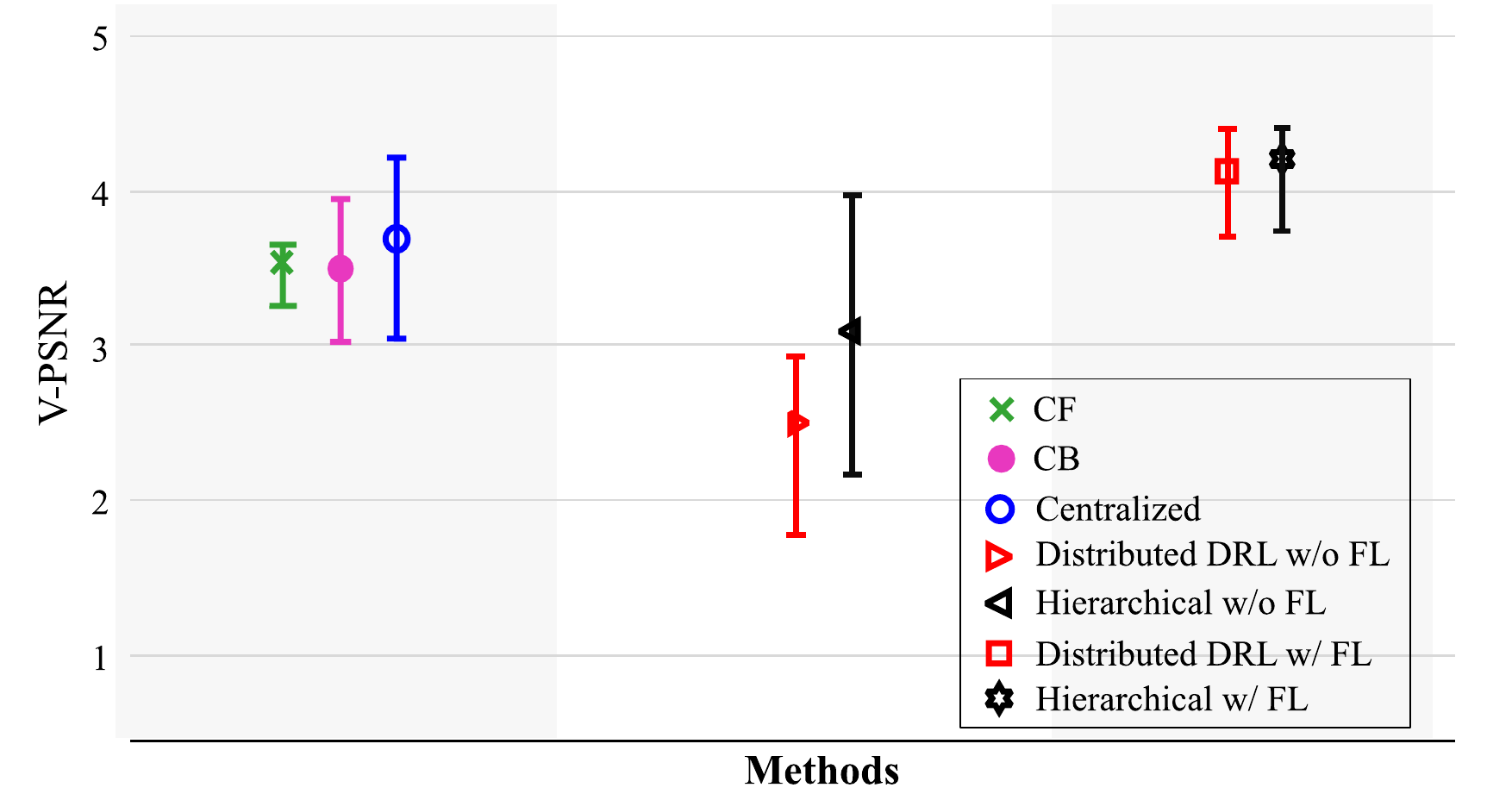}
    \caption{V-PSNR Performance of proposing algorithms.}
    \label{fig:performance}
    \end{minipage}
\end{figure}

To ease the presentation of V-PSNR, we normalize the resulting V-PSNR value into $[0,5]$ (5 frames in each GOP). Note that, the DRL algorithm is well-known for its lack of reliability. Average performance is not sufficient to describe the performance of the algorithm. To show the risk of our algorithm \cite{Chan2019}, we use a standard derivative (SD) error bar to show the performance. We present +std, average performance, and -std, V-PSNR value over $10^5$ random GOPs with independently generated UAV and VR users. For each algorithm setting, we train $6\times 10^4$ epochs and pick the best model during training to plot the result. In the following, we use "Centralized(Reduced)", "Distributed DRL w/ FL", and "Hierarchical w/ FL" to denote the centralized DRL association algorithm with P-PF scheduler, federated distributed DRL algorithm with P-PF scheduler, a hierarchical algorithm with federated distributed DRL and learning-based scheduler algorithm, respectively. To show the effectiveness of FL, we compare two more algorithms: Distributed DRL without FL and Hierarchical FL without FL. For simplicity, we use "Distributed DRL w/o FL" and "Hierarchical FL w/o FL", respectively.

\subsection{Overall Convergence and Policy Visualization}
Fig. \ref{fig:psnr} plots the overall V-PSNR versus the training epochs.
In Fig. \ref{fig:psnr_3}, we observe that the Centralized(Reduced), Distributed DRL w/ FL and Hierarchical w/ FL converge fast within 10,000 epochs. Because the FL method combines the knowledge among APs. In Fig. \ref{fig:psnr_33}, we observe that the Distributed DRL w/o FL fails to converge, whereas the Hierarchical w/o FL approach does converge but slower. Because the Distributed DRL w/o FL approach treats other agents with dynamic policies as part of the environment. Reinforcement learning does not have a convergence guarantee for a non-stationary environment. Federated learning can reduce the variance of the environment \cite{8792117}. In the Hierarchical w/ FL approach, the agent of scheduler acts as a meta-controller, who helps the distributed DRL agents to cooperate. Fig. \ref{fig:performance} plots the overall V-PSNR values of different learning algorithms. we observe that the average V-PSNR of algorithms follows: Hierarchical w/ FL $\approx$ Distributed DRL w/ FL $>$ CB $\approx$ CF $>$ Hierarchical w/o FL $>$ Distributed DRL w/o FL.

\begin{figure*}[t]
    \centering
    \includegraphics[width = 0.85 \textwidth]{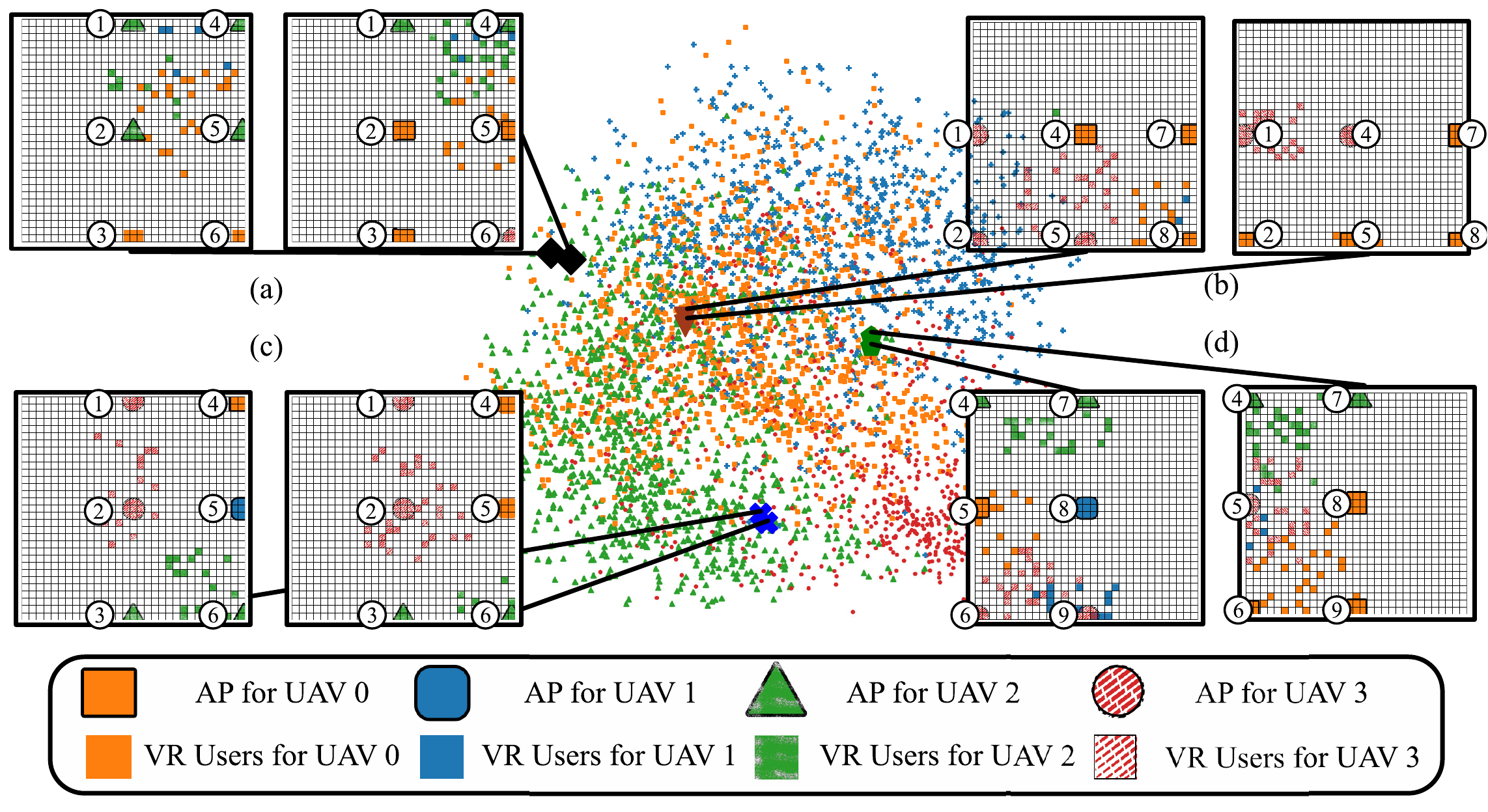}
    \caption{t-SNE embedding of the representations with the learned policy.}
    \label{fig:tsne}
\end{figure*}



In Fig. \ref{fig:tsne}, we show the generated hidden vectors of our proposed neural network by visualising the output of CNN and the final policy. The figure is generated with $10^4$ randomly generated environment examples. With generated vectors from the output of CNN layers, we apply a technique developed for the visualization of high-dimensional data called \say{t-SNE} to calculate the distance between vectors. Then, the principle composition analyses (PCA) is performed on the vectors to reduce the dimension to 2D space and visualise them in Fig. \ref{fig:tsne}. Each point is coloured according to the association decisions \cite{Mnih2015}. In this way, the point cloud presents how the neural network recognises the environment and makes decisions.

In Fig. \ref{fig:tsne}, we also randomly present four pairs of observations together with their most similar observations by picking the nearest one according to the result of the t-SNE algorithm. Each observation is observed by the AP in the center, which is represented as a grid-map. In each grid-map, the colours in grids represent the position of VR users and their corresponding UAV. The position of APs is also marked and coloured by its association decision from current observation. To ease the reading of these figures, we number the $9$ APs in our simulation with numbers from $1-9$ based on their relative positions. In (a), the grid-maps are observed by $2$nd AP. In left grid-map, we can see that $2$nd and $5$th AP is jointly associated to serve $2$nd UAV. In right grid-map, $2$nd, $5$th APs jointly serve $0$th UAV. In (b), the $4$th APs forms virtual cells with $7$th and $8$th APs cooperatively to serve $0$th UAV in the left grid-map. In right grid-map, $4$th and $1$st APs jointly serve $3$th UAVs. In (c), the $1$st and $2$nd APs jointly serve the surrounding VR user groups, which request tiles from $3$rd UAV. In (d), the left observation from $8$th AP shows that it fails to cooperate with $9$th AP to serve $2$nd UAV and corresponding VR user group. This highlights the fact that the value of the actions in each agent is jointly decided by its and its neighbours' actions in the multi-agent system. The right observation has shown that $6$th, $8$th, and $9$th APs jointly serve $0$th UAV, whereas the $4$th and $7$th APs jointly serve $2$th UAV.

\subsection{Quality-of-experience Analysis}
In this subsection, we plot the V-PSNR value using VR users of three learning algorithms, including \textcolor{black}{Centralized(Reduced), Distributed DRL w/ FL, and Hierarchical w/ FL,} together with two conventional algorithms (CB, CF) in different scenarios. We show the generalization and effectiveness of our proposing algorithms.

\begin{figure}[t]
    \begin{minipage}[t]{0.48\textwidth}
    \centering
    \includegraphics[width = \textwidth]{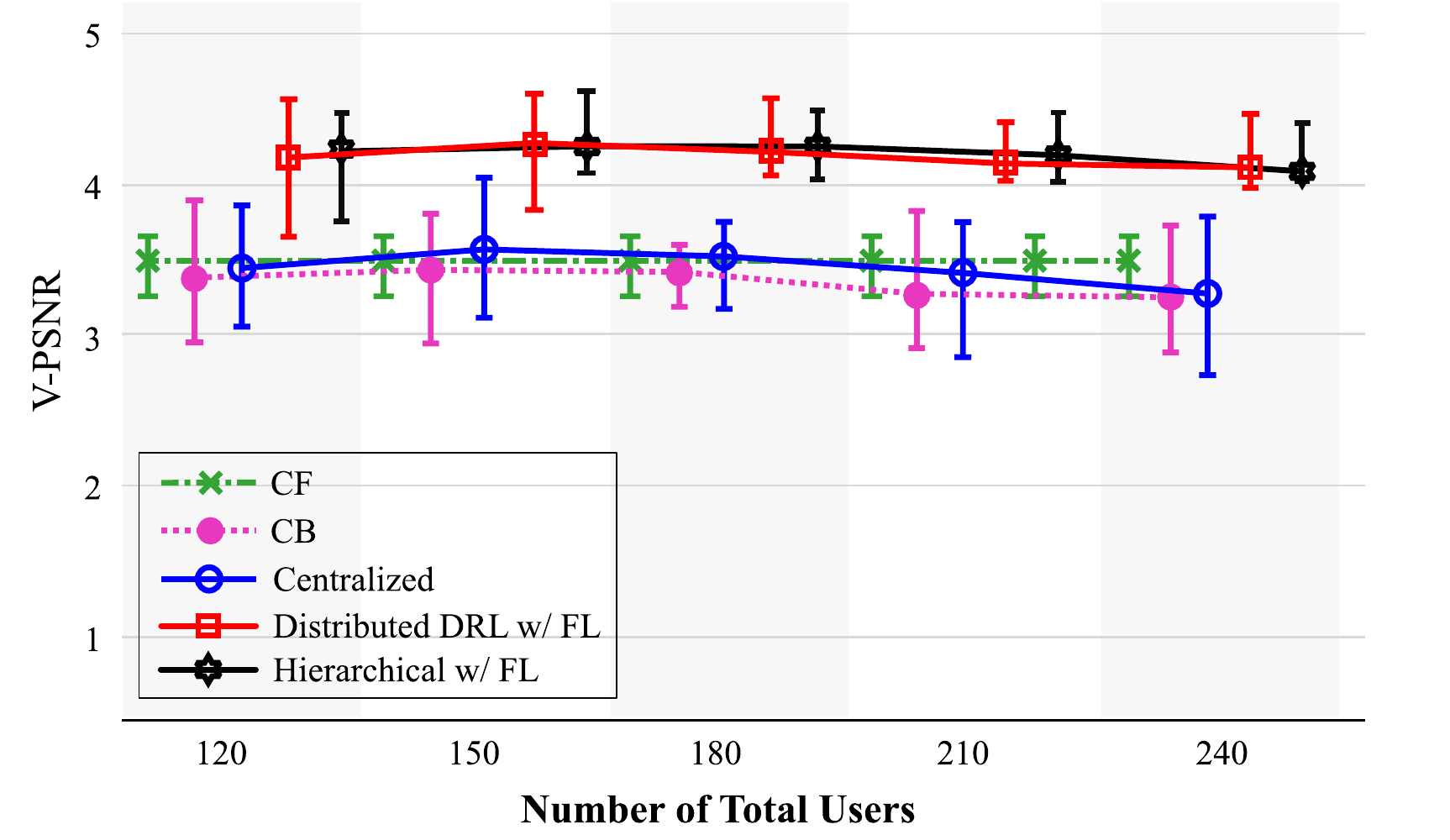}
    \caption{V-PSNR of our proposing algorithms with different number of VR users.}
    \label{fig:psnr-numofuser}
    \end{minipage}
    \hspace*{0.4cm}
    \begin{minipage}[t]{0.48\textwidth}
    \centering
    \includegraphics[width = \textwidth]{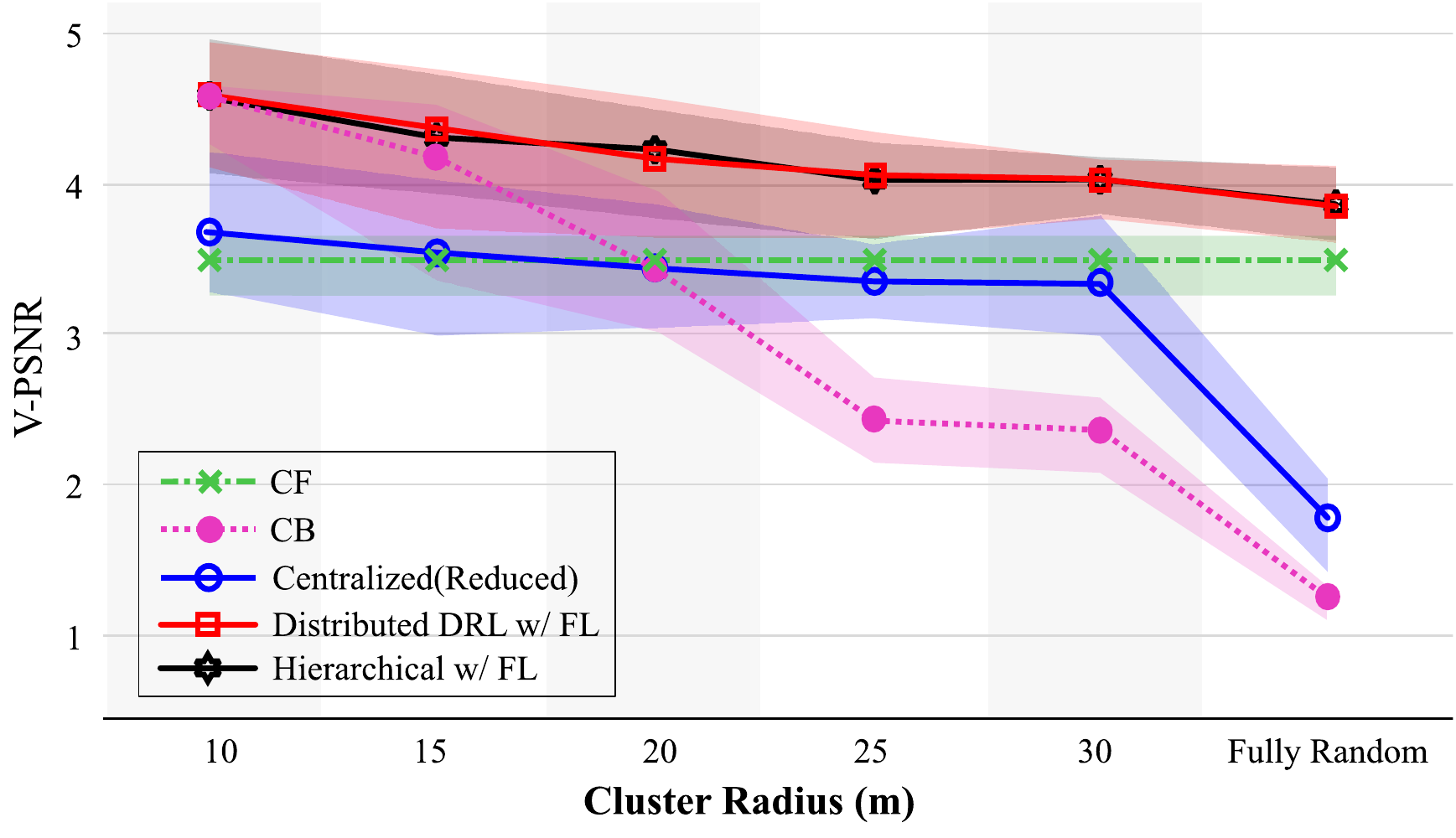}
    \caption{V-PSNR of our proposing algorithms with different VR users' cluster radius.}
    \label{fig:psnr-radius}
    \end{minipage}
\end{figure}

Fig. \ref{fig:psnr-numofuser} plots the V-PSNR value versus the number of VR users. We observe that all algorithms' V-PSNR stay nearly unchanged with increasing numbers of VR users in CF-MB network. This matches our expectation for CF-MB network, where the UAV-APs cooperative reception enhance the received signal from the UAV and the APs-VR broadcasting is not sensitive to the number of receiving VR users. It is worth mentioning that we only train a single model using random VR users and obtain similar results with different numbers of VR users setting.

Fig. \ref{fig:psnr-radius} plots the V-PSNR value versus the VR users' cluster radius. We observe that the V-PSNR value of our proposed distributed algorithms, including Distributed DRL w FL and Hierarchical w FL, drop slightly with the increasing cluster radius, but outperform other algorithms (CF, CB). Because the reuse of frequency resources spatially improves the performance. The V-PSNR value of CF association algorithm keeps the same for different cluster radius, as all APs jointly serve one UAV and corresponding VR user group in the CF association. \textcolor{black}{The CF-MB network provides uniform services in this case, which lacks geographical awareness and won't work in a large-scale network.}
We observe that the V-PSNR of the CB association algorithm drops dramatically with the increasing cluster radius. The reason is that the increasing cluster radius can lead to more overlap clusters, high inter-cell interference and poor cell-edge performance.

\begin{figure}[t]
    \begin{minipage}[t]{0.48\textwidth}
    \centering
    \includegraphics[width = \textwidth]{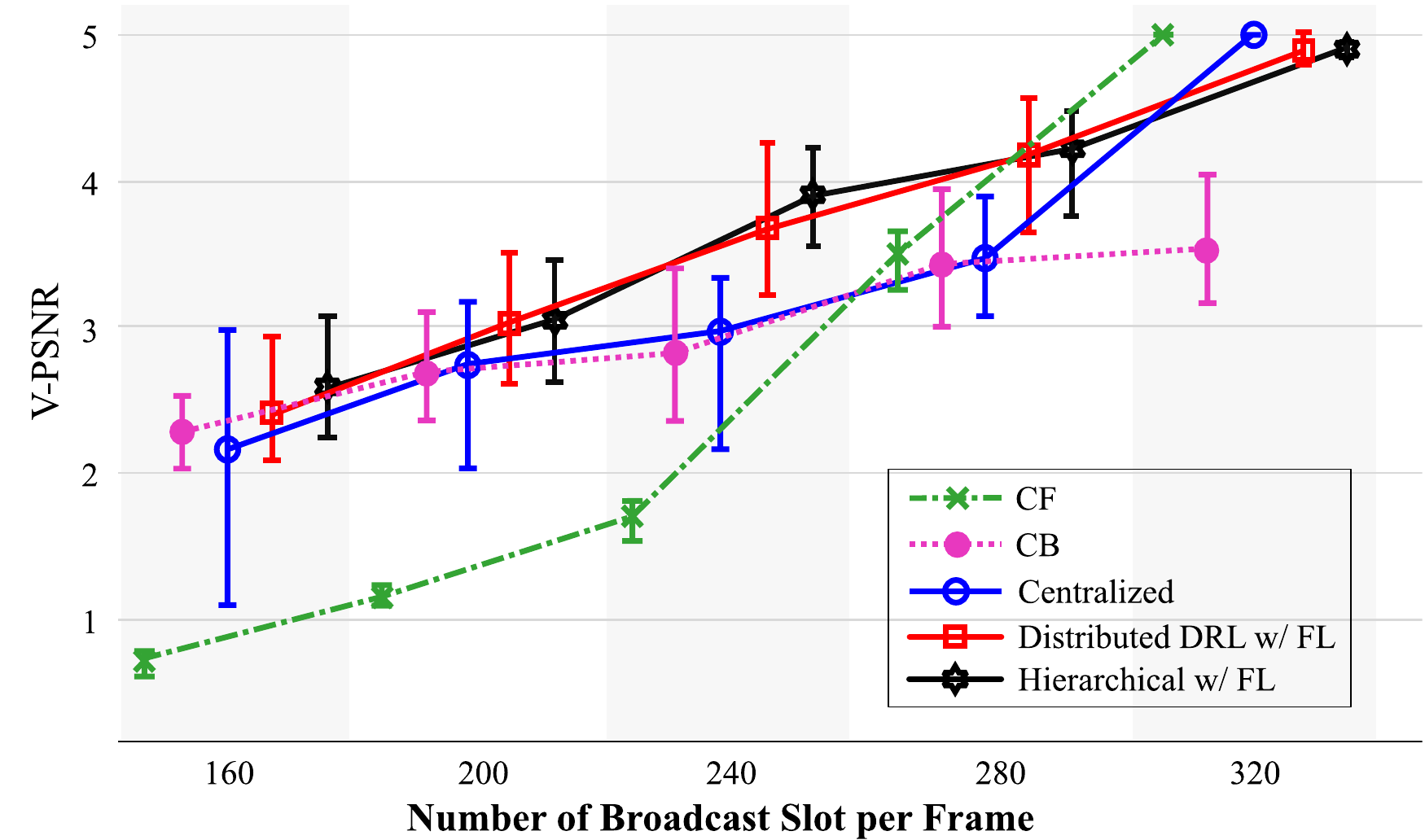}
    \caption{V-PSNR of our proposing algorithms with different broadcast slots in each frame duration.}
    \label{fig:psnr-resource}
    \end{minipage}
    \hspace*{0.4cm}
    \begin{minipage}[t]{0.48\textwidth}
    \centering
    \includegraphics[width = \textwidth]{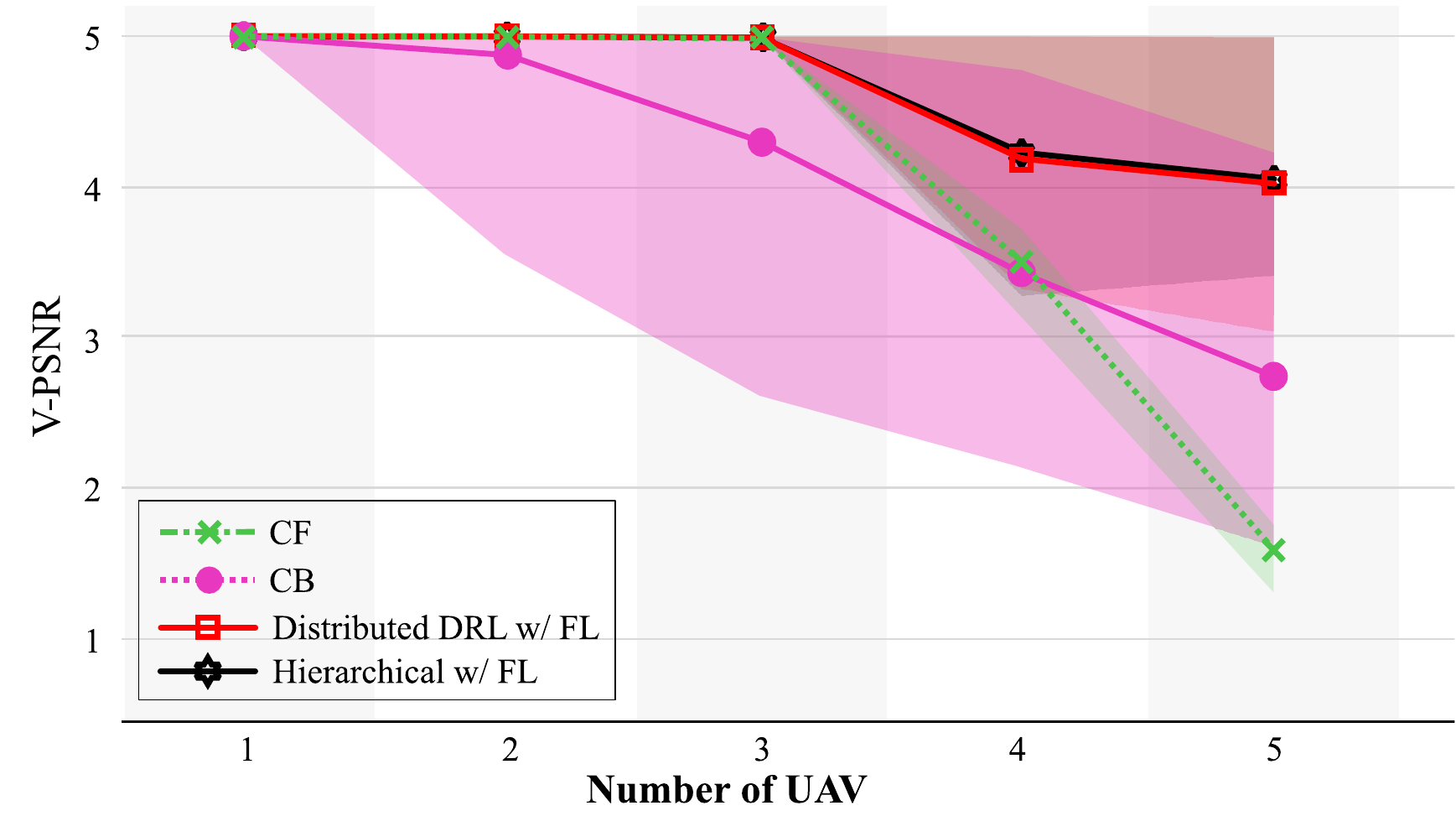}
    \caption{V-PSNR of our proposing algorithms with different number of UAVs.}
    \label{fig:psnr-uav}
    \end{minipage}
\end{figure}

Fig. \ref{fig:psnr-resource} plots the V-PSNR versus the number of broadcast slots, which also reveals the slot utilization of our proposed algorithms. Remind that in our considered environment, 4 UAV holds 288 tiles in total. If we set large $T_\text{b}$ and fewer broadcast slots, then two tiles should be fully transmitted successfully within one broadcast slot (160 slots). If we set small $T_\text{b}$ (more broadcast slots), each tile can occupy one broadcast slot individually (320 slots).
We observe that the V-PSNR of CF association method increases with the number of broadcast slots, as no interference in each slot. The V-PSNR value of CB approaches decrease with the increase of broadcast slots with lower per-slot utilization. By dynamically arranging the association policy, our proposed algorithm can always achieve the maximum slot utility among different approaches.

Fig. \ref{fig:psnr-uav} plot the V-PSNR value of different algorithms versus the increasing number of UAVs. We observe that the V-PSNR of CF and CB decreases with the increasing number of UAVs due to the lack of resources. We can see that the learning-based algorithms still outperform conventional methods. \textcolor{black}{It achieves high utilization for each broadcast slot with increased UAV number in both average and standard derivation of V-PSNR. It should be noted that the training complexity of learning algorithms increases linearly with the increasing number of UAV in our network design, which becomes the most important factor limiting the scalability of our algorithm.}

\section{Conclusion}
In this paper, we introduced a cell-free multi-group broadcast network for real-time VR video transmission from UAVs to VR users for experience enhancement in a sports event. 
To optimise the quality-of-experience of VR users with dependent decoded video resources and correlated VR users, we highlighted the importance of scheduling video tiles and the dynamical association of APs. We have also shown that a joint design is needed for correlated and sequential scheduling and association procedures. To explore the learning-based dynamic association algorithm, we propose a centralized and multi-agent deep reinforcement learning algorithm, which captures the environment via convolutional layers. To jointly solve the coupled association and scheduling algorithm, we further developed a hierarchical algorithm with scheduler as meta-controller and association algorithm as the controller.
Our results demonstrated that both distributed APs and hierarchical with federated learning algorithms can effectively handle a large number of APs and VR users and outperform the centralized algorithm and non-learning-based approach with decent scalability.


\end{document}